\documentclass[reprint,floatfix,notitlepage,nofootinbib,twocolumn,superscriptaddress,prb]{revtex4-1}
\usepackage{amsmath}
\usepackage{amssymb}
\usepackage{graphicx}
\usepackage[tight]{subfigure}
\usepackage{enumitem}
\usepackage{soul}
\usepackage{cancel}
\usepackage{tikz}
\usepackage{pifont}
\usepackage{xspace}
\usepackage{comment}

\makeatletter
\renewcommand{\p@subsection}{}
\renewcommand{\p@subsubsection}{}
\makeatother

\usepackage[english]{babel}
\makeatletter
\def\bbl@set@language#1{%
  \edef\languagename{%
    \ifnum\escapechar=\expandafter`\string#1\@empty
    \else\string#1\@empty\fi}%
  \@ifundefined{babel@language@alias@\languagename}{}{%
    \edef\languagename{\@nameuse{babel@language@alias@\languagename}}%
  }%
  \select@language{\languagename}%
  \expandafter\ifx\csname date\languagename\endcsname\relax\else
    \if@filesw
      \protected@write\@auxout{}{\string\select@language{\languagename}}%
      \bbl@for\bbl@tempa\BabelContentsFiles{%
        \addtocontents{\bbl@tempa}{\xstring\select@language{\languagename}}}%
      \bbl@usehooks{write}{}%
    \fi
  \fi}
\newcommand{\DeclareLanguageAlias}[2]{%
  \global\@namedef{babel@language@alias@#1}{#2}%
}
\makeatother
\DeclareLanguageAlias{en}{english}

\usetikzlibrary{arrows}
\usetikzlibrary{intersections}
\usetikzlibrary{shapes.geometric}
\usetikzlibrary{decorations.pathmorphing, patterns,shapes,fixedpointarithmetic}
\usetikzlibrary{decorations.markings}

\tikzset{
  mid arrow/.style={postaction={decorate,decoration={
        markings,
        mark=at position .575 with {\arrow{stealth}}
      }}},
  end arrow/.style={postaction={decorate,decoration={
        markings,
        mark=at position 1 with {\arrow{stealth}}
      }}},
  snake arrow/.style={fixed point arithmetic, decorate, decoration={snake,amplitude=2pt, segment length=11pt},postaction={decoration={markings,mark=at position 0.625 with {\arrow{stealth}}},decorate}},
}

\usepackage{bm}

\newcommand{\ket}[1]{|#1 \rangle}
\newcommand{\bra}[1]{\langle #1|}
\usepackage[pdfusetitle,colorlinks=true,citecolor=blue,linkcolor=magenta]{hyperref}

\graphicspath{{./}{./images/}}
\begin{document}
\title{Multifractality in non-unitary random dynamics}

\author{Jason Iaconis}
\affiliation{Department of Physics and Center for Theory of Quantum Matter, University of Colorado, Boulder CO 80309, USA}

\author{Xiao Chen}
\affiliation{Department of Physics, Boston College, Chestnut Hill, MA 02467, USA}

\begin{abstract}
 We explore the multifractality of the steady state wave function in non-unitary random quantum  dynamics in one dimension. We focus on two classes of random systems: the hybrid Clifford circuit model and the non-unitary free fermion dynamics. In the hybrid Clifford model, we map the measurement driven transition to an Anderson localization transition in an effective graph space by using properties of the stabilizer state. We show that the volume law phase with nonzero measurement rate is non-ergodic in the graph space and exhibits weak multifractal behavior. We apply the same method to the hybrid Clifford quantum automaton circuit and obtain similar multifractality in the volume law phase. For the non-unitary random free fermion system with a critical steady state, we compute the moments of the probability distribution of the single particle wave function and demonstrate that it is also weakly multifractal and has strong variations in real space.
 
\end{abstract}

\maketitle

\section{Introduction}
Non-unitary dynamics have attracted a lot of attention in the past few years. It has been shown that for a generic many-body unitary quantum  dynamics subject to local projective measurement, there exists an entanglement phase transition at the level of the quantum trajectories \cite{li2018quantum,  skinner2019measurement, gullans2020dynamical,chan2019unitary,zabalo2020critical, gullans2020scalable, Li_2019, iaconis2020measurement,   jian2020measurement, bao2020theory,Tang_Zhu_2020}. In the steady state wave function, by increasing the measurement rate $p$, the entanglement entropy changes from a highly entangled volume-law scaling to a short-range entangled area-law scaling. In particular, when $p$ is nonzero, the volume-law phase has a non-trivial subleading correction term which is absent in the conventional thermal phase \cite{Li_2019, gullans2020dynamical,fan2020self,li2020statistical}. The stability of this non-thermal volume law phase has interesting interpretations in the language of quantum error correction \cite{choi2020quantum, gullans2020dynamical,fan2020self,li2020statistical}. 



For this measurement induced phase transition, the dynamics is random in both space and time. The randomness comes from various sources including the choice of unitary gate, the position of the measurement gate and the measurement outcome. The presence of the randomness leads to an emergent critical point which is distinct from any conventional critical point in a clean system\cite{skinner2019measurement,li2020conformal}. Furthermore, in the highly entangled non-thermal volume law phase, a subleading correction to the entanglement entropy is caused by the fluctuation of the random dynamics and is different from the prediction given by simple mean-field theory estimation \cite{li2020statistical,fan2020self,li2021entanglement}.   This observation implies the random fluctuation effect becomes dominant in the low dimensional quantum dynamics and renders the physics significantly different from that in the clean systems.

In this paper, we will go beyond this entanglement picture and investigate multifractal behavior in the non-unitary random dynamics.  Multifractality has been observed in many random systems and historically has played an important role in identifying the Anderson localization \cite{Wegner,Evers_RMP} and spin glass phase transitions \cite{Derrida_REM}. For instance, the Anderson localization phase transition in disordered systems is a continuous phase transition separating the extensive metallic state from a localized state. At the critical point, the single particle wave function has strong spatial fluctuations and is multifractal in nature. This can be characterized by using the inverse participation ratio (IPR): 
\begin{align}
    I_q=\int d^d{\bf r} |\psi({\bf r})|^{2q}\sim L^{-\tau_q},
\end{align}
where $\psi({\bf r})$ is the normalized single particle wave function in real space and the exponent $\tau_q$ is an infinite set of critical exponents describing the moments of $|\psi({\bf r})|^2$ \cite{Wegner,Evers_RMP}.
We can further introduce the fractional dimension $D_q$ via the relation $\tau_q=D_q(q-1)$, in order to quantify how extended the wave function is. At the critical point,  $D_q$ takes a fractal value and has a non-trivial dependence on $q$. In contrast, in the metallic phase, the wave function is uniform in the space with $D_q=d$ (the spatial dimension of the system), while in the localized phase, the wave function is exponentially localized with $D_q=0$.

In this paper, we first consider hybrid random Clifford circuits in which the wave functions can be represented using the stabilizer formalism \cite{Li_2018,Li_2019}.  Using this formalism, the steady state wave functions can be transformed into a so called graph state by applying only local unitary operations. We analyse the IPR of the eigenvectors of the adjacency matrix associated with this graph. We find that throughout the non-thermal volume law phase, the graph has a high connectivity and $D_q$ takes a fractional value between 0 and 1. Furthermore, this fractional dimension, $D_q$, has a non-trivial dependence on $q$, demonstrating the multifractal nature of these graph states. When the measurement rate $p>p_c$, the random graph obtains a local structure with low connectivity and has $D_q=0$, analogous to the Anderson localized phase. We further apply the same method to the hybrid Clifford quantum automaton circuit \cite{iaconis2020measurement} and find similar multifractal behavior in the volume law phase. 

In addition, we consider non-unitary random free fermion dynamics in one dimension. Previous studies indicate the existence of a critical phase in this model, which enjoys emergent two dimensional conformal symmetry with a spacelike time direction \cite{Chen_2020}. We observe that in the critical steady state, the single particle wave function has strong fluctuations in space and is multifractal in nature. This provides strong evidence that the wave function is qualitatively different from the critical state of the clean free fermion system, in which the single particle wave function's amplitude is uniform in space and the criticality comes from quantum coherence effects.

The rest of the paper is organized as follows. In Sec.~\ref{sec:hybrid_clifford}, we study the multifractal behavior in the hybrid random Clifford circuit. We first review both the stabilizer and the graph state formalism in Sec.~\ref{sec:stabilizer}. We then compute the IPR of the corresponding graph state in the steady state of the hybrid random Clifford circuit model in Sec.~\ref{sec:adjacency}. We apply the same method to the hybrid Clifford quantum automaton circuit in Sec.~\ref{sec:automaton}. In Sec.~\ref{sec:fermion}, we analyze the multifractal behavior of the non-unitary free fermion dynamics. In Sec.~\ref{sec:conclusion} , we summarize our results and discuss possible directions for future work.

\section{hybrid random Clifford circuit}
\label{sec:hybrid_clifford}
In this section, we study the dynamics of hybrid quantum circuits which are composed solely of 2-site random unitary gates drawn uniformly from the Clifford group interspersed with a layer of projective Pauli measurements (See Fig.~\ref{fig:cartoon_clifford}). This model exhibits an entanglement phase transition from a volume law phase to an area law phase as we vary the measurement rate $p$ \cite{}. Using the stabilizer formalism, it is possible to efficiently simulate this Clifford dynamics and analyze the scaling of the entanglement in both of these phases as well as at the critical point $p_c$. In this work, we study the so called graph states, which can always be obtained from the steady state stabilizer wave function \cite{Raussendorf_2003,Van_den_Nest_2004}. We will show that this ensemble of random graphs, which can be specified by a corresponding adjacency matrix, possess eigenstates which exhibit a localization transition at $p_c$. Furthermore, we will show that in the regime $0<p<p_c$, these eigenstates  show the characteristic properties of multifractality.

\subsection{Stabilizer formalism}
\label{sec:stabilizer}
In Clifford circuits, the dynamics are conveniently described using the notion of stabilizer operators. These stabilizers can be used to completely define a class of quantum states and simulate their quantum dynamics. In this subsection, we give a brief review of the stabilizer formalism \cite{gottesman1998heisenberg,Aaronson_2004}.

\begin{figure}
    \centering
     \includegraphics[scale=0.35]{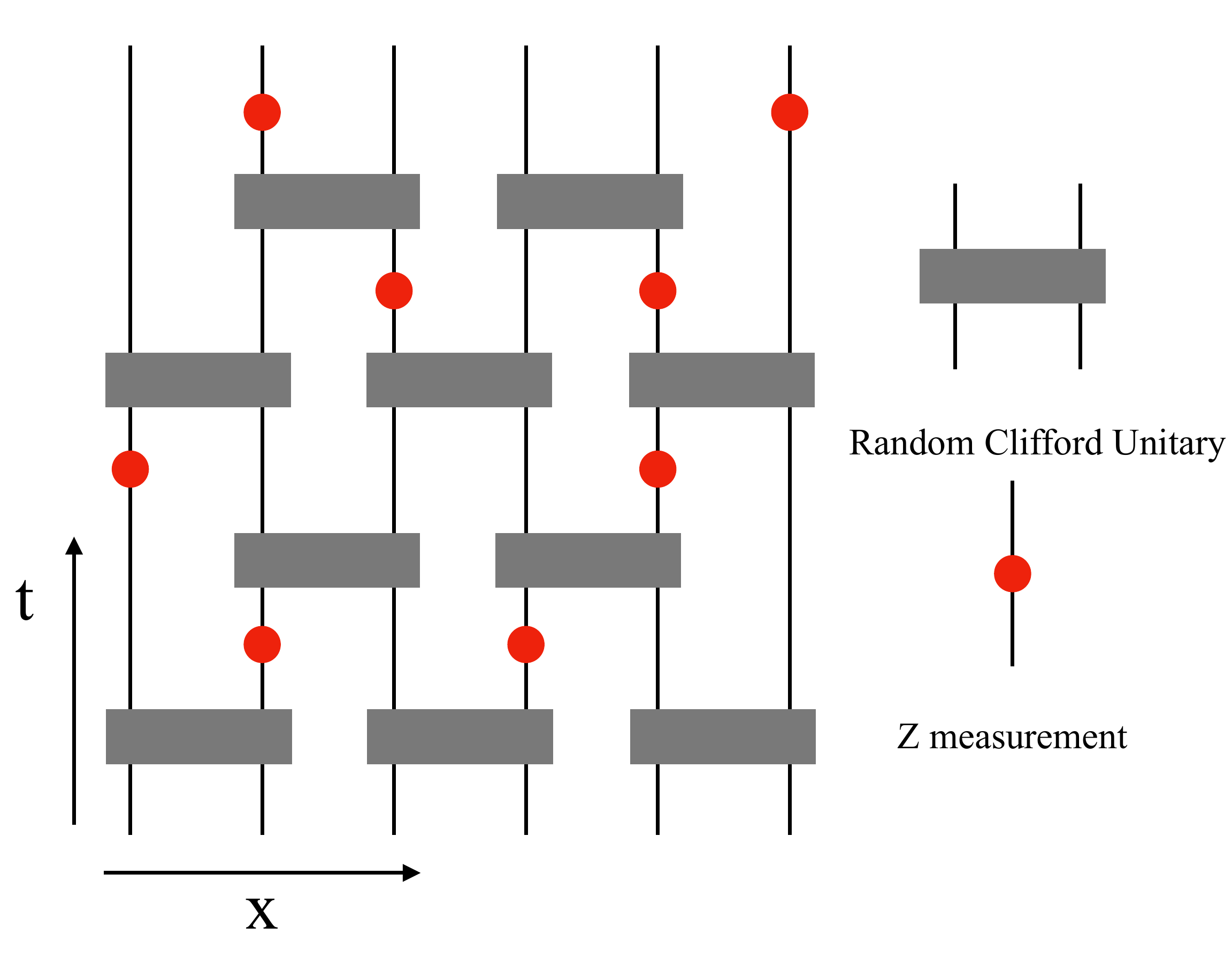}
    \caption{The cartoon for the one dimensional hybrid random Clifford circuit. The random two-qubit unitary gates are arranged in a brick-wall fashion, while the single-qubit Z measurements are  randomly applied with probability $p$ at each time.}
    \label{fig:cartoon_clifford}
\end{figure}

An $L$-qubit stabilizer state, $|\psi\rangle$, is completely defined as the simultaneous eigenstate of $L$ commuting
and independent Pauli string operators $M_i$ with eigenvalue +1.  The $M_i$ operators are the generators of the stabilizer group $\mathcal{S}$, which is a subgroup of the $L$-qubit Pauli group. We can write each generator as 
 $M_i=X_1^{a^i_1}Z_1^{b^i_1}X_2^{a^i_2}Z_2^{b^i_2}\cdots X_L^{a^i_L}Z_L^{b^i_L}$, with $a^i_n,b^i_n$ taking values 0 or 1. The information contained in these $L$ vectors, $(a^i_1,a^i_2,\cdots,a^i_L,b^i_1,b^i_2,\cdots,b^i_L)$, can be arranged into the so called ``stabilizer tableau". This is a $L\times 2 L$ binary matrix $T=[T_X,T_Z]$, where the first square matrix, $T_X$, stores the information of $\{a\}$ and the second matrix, $T_Z$, stores the information contained in $\{b\}$.  For instance, a trivial product state in the z-direction $|\psi\rangle=|00\cdots 0\rangle$ has $M_i=Z_i$ for all $i=1\dots L$. In the corresponding stabilizer tableau, $T_X$ is a zero matrix and $T_Z$ is the identity matrix. Under the hybrid Clifford dynamics described in Fig.~\ref{fig:cartoon_clifford}), the stabilizer operators, $M_i$, are mapped to a new set of stabilizer operators $M_i^\prime$. Therefore, the wave function remains a stabilizer state, with the stabilizer tableau being updated accordingly
 \footnote{We disregard the phase information since it is not important in our paper.}. 
 
The R\'enyi entanglement entropy for a subsystem A is defined as $
S_n=\frac{1}{1-n}\log_2\mbox{Tr}\rho_A^n$ where $\rho_A$ is the reduced density matrix for region A.
For the stabilizer states, $S_n$ is independent of the R\'enyi index $n$ and obeys the form \cite{Hamma}
\begin{align}
    S_n=L_A-\log_2|\mathcal{S}_A|,
\end{align}
where $\mathcal{S}_A$ is a stabilizer group defined in subsystem A, which is a subgroup of $\mathcal{S}$.  $|\mathcal{S}_A|$ counts the number of independent stabilizers supported only on A.

In the hybrid random Clifford dynamics, when the measurement rate $p$ is small, the steady state entanglement entropy has volume law scaling. Roughly speaking, this implies that the corresponding stabilizer generators of this state  span the entire system. A single projective measurement replaces one stabilizer with a local operator. As we increase $p$, the long stabilizers are gradually replaced by a series of short stabilizers.  Eventually, at a high enough measurement rate $p>p_c$, the steady state entanglement entropy has area law scaling.

\begin{figure*}
    \centering
    \includegraphics[scale=0.5]{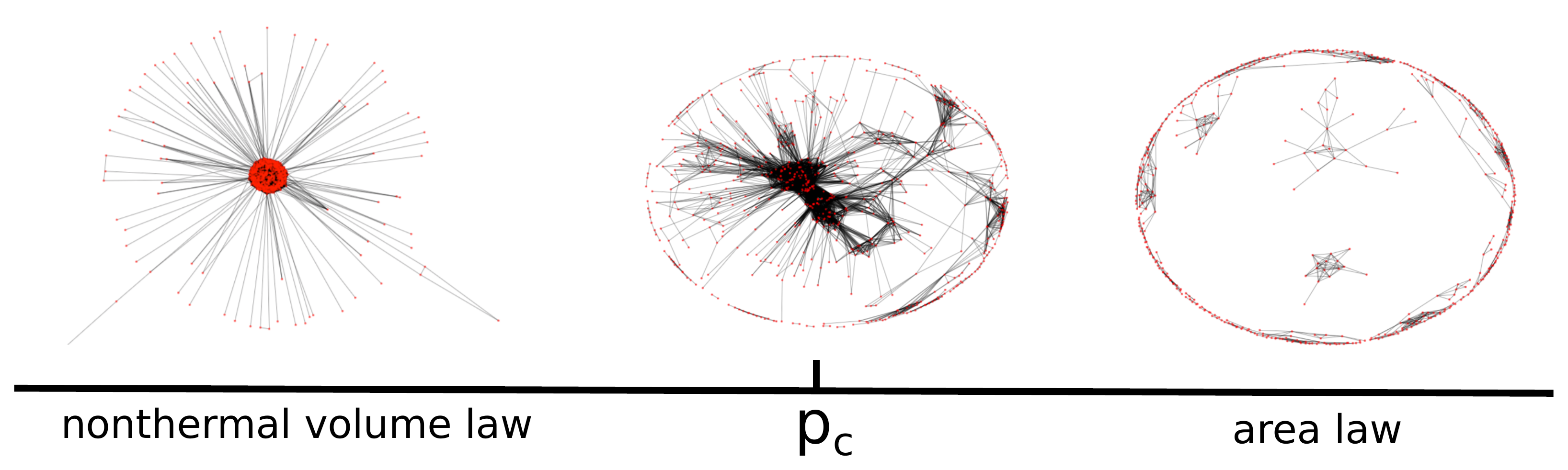}
    \caption{The emergent graph state which results from the random Clifford dynamics with a measurement rate corresponding to the {\it (left)} volume law, {\it (middle)} critical and {\it (right)} area law entangled phases. Each node of the graph represents a single qubit, with the edges corresponding to the associated CZ gates which are applied to a reference state to create the graph state. The placement of the nodes are determined according to the Fruchterman-Reingold force-directed drawing algorithm which treats each node as a particle and introduces forces between them \cite{fruchterman1991graph}. There are attractive forces between adjacent nodes and repulsive forces between all pairs of nodes. The position of the nodes is fixed by minimizing the energy of the entire system. In the non-thermal volume law phase, a finite fraction of the nodes are connected with $O(L)$ other nodes and forms the ``core" of the graph. The outer layers of the graph is composed of the nodes which are connected with only a few number of other nodes. As we increase the measurement rate $p$, the core of the graph becomes smaller and more nodes are pushed to the outer layers of the graph. At the critical point, the core disappears and the graph has a self-similar structure. In the area law phase, most of the nodes are living on the outermost circle and are only connecting with nearby nodes. }
    \label{fig:random_graphs}
\end{figure*}

\subsubsection{Graph states}

In this subsection, we introduce an important subclass of stabilizer states known as graph states. We briefly review the definition of the graph states and their connection with the more general stabilizer states.

An undirected graph, $G$, is defined by a set of vertices $V$ and edges $E$. For any graph, $G$, we can define a corresponding graph state $\ket{\psi(G)}$ as
\begin{eqnarray}
\ket{\psi} = \prod_{(ij) \in E} \Lambda_{ij}(Z)\ket{+}^{\otimes |V|}
\end{eqnarray}
where  $\Lambda(Z)_{ij} = {\rm I}-2\ket{11}\bra{11}$ is the two-qubit control Z (CZ) unitary operator and $\ket{+}^{\otimes |V|}$ is the reference state with  $|+\rangle=(|0\rangle+|1\rangle)/\sqrt{2}$ polarized in x direction.
 The stabilizer generators of such a graph state are given by the set
\begin{eqnarray}
M_i = X_i \prod_{j | (ij) \in E} Z_j.
\end{eqnarray}
In the language of the stabilizer tableau, $T_X$ is an identity matrix. Since all $M_i$ operators commute with each other, $T_Z$ is required to be a symmetric binary matrix. We further require that all the diagonal elements of $T_Z$ are zeros. This  Z stabilizer tableau is exactly the adjacency matrix of the graph $G$. When $T^{ij}_Z=1$, we have $(i,j)\in E$ and $T^{ij}_Z=0$ otherwise. 

A key feature for our analysis is that all stabilizer states are equivalent to a graph state up to the application of {\it single qubit} unitary rotations S (phase gate) or H (Hadamard gate) \cite{Van_den_Nest_2004}. In the language of the stabilizer tableau, this mapping onto a graph state can be done in two steps: (1) We swap/add the rows of $T$ (Gaussian elimination over finite $\mathbb{Z}_2$ field) to transform $T_X$ into an upper triangular matrix. In this process, we may also apply H gates to enforce that all the diagonal elements of $T_X$ are equal to one. (2) We add rows in $T$ to transform $T_X$ into an identity matrix. We further apply S gates to enforce the condition that $T_Z$ contains only zeros along its diagonal. Note that row operations which are applied to the stabilizer tableaux do not change the stabilizer wave function. Since the only nontrivial operations we apply are single qubit unitary gates, the entanglement entropy is invariant under this transformation.

As a consequence, for a stabilizer state evolved under the hybrid Clifford dynamics, at any time, it can be transformed into a graph state with the complete quantum information contained in the corresponding adjacency matrix. Its entanglement entropy is closely related to the connectivity properties of the underlying graph. In particular, if we bipartition a graph into two subsets A and $\overline{\rm A}$ with
\begin{align}
    T_Z=\begin{pmatrix}
    T_Z^{AA} & T_Z^{A\bar{A}}\\
        T_Z^{\bar{A}A} & T_Z^{\bar{A}\bar{A}}
    \end{pmatrix},
\end{align}
the connectivity between A and $\overline{\rm A}$ can be quantified by ${\rm rank}_2(T_Z^{A\bar{A}})$, which is exactly the entanglement entropy of A \cite{hein2006entanglement}. Stabilizer wave functions in the volume law phase are characterized by graphs with very high connectivity which lack locality between the connected vertices of the graph.  On the other hand, in the area law phase, the adjacency matrix is more sparse and  vertices are only connected to other vertices which are spatially nearby in the original circuit construction.  

In the following section, we will show that there exists a structural change of the adjacency matrix across $p_c$ which is similar to the Anderson localization transition of random matrix models. One of the most prominent examples of such a model is the power-law random banded matrix ensemble, in which the off-diagonal elements of a random matrix have zero mean and variance $1/r^{2\alpha}$, where $r$ is the distance from the diagonal element \cite{Evers_PRL,Evers_RMP}. This matrix describes a one dimensional random free fermion  Hamiltonian with long-range hopping. As we increase $\alpha$, the hopping becomes increasingly local and the single particle eigenstate undergoes an Anderson localization transition from an extended state to a localized state. Precisely at the transition point $\alpha=1$, the  critical eigenstate is neither localized or extended. The calculation of moments indicates that the eigenstate has strong fluctuations and is multifractal -- characterized by a infinite set of fractal dimensions \cite{Evers_RMP}. Motivated by these studies, we map the hybrid circuit measurement driven transition to an Anderson localization transition in the effective graph space. We diagonalize the adjacency matrix and analyze the possible multifractal behavior in its eigenstates.

In Fig.~\ref{fig:random_graphs}, we show three examples of characteristic graphs which describe the random Clifford circuit stabilizer states $i)$ deep within the volume phase $0<p<p_c$, $ii)$ at the critical point $p_c$, and $iii)$ in the area law phase $p>p_c$. In the volume law phase, when there is a nonzero measurement rate, the graph consists of a core of very highly connected nodes plus an outer shell of nodes with low connectivity. The highly connected core leads to volume law entanglement scaling and the same structure has also been observed in the pure unitary evolution with $p=0$. The presence of the outer shell nodes is due to the measurement and as we will show later, gives rise to a multifractal structure. At exactly the critical point $p=p_c$, the inner core disappears. The graph at this point has a complex self similar structure and the entanglement entropy scales logarithmically in the subsystem size.  Finally in the area law phase, the connectivity  is dramatically reduced and the graph has an emergent local structure.  In the following subsection, in order to quantitatively characterize the structure of these graphs, we will analyze the properties of the eigenvectors and eigenvalues of the adjacency matrix.

\subsection{Adjacency matrix and multifractality}
\label{sec:adjacency}
As we discussed before, each graph state is specified by an adjacency matrix $T_Z$. Since it is a hermitian binary matrix, we can treat it as a Hamiltonian for a free fermion system with 
\begin{align}
    H=\sum_{ij}T_Z^{ij}c_i^\dag c_j.
\end{align}
This model describes a free fermion hopping on an ensemble of random graphs, $G$, generated by the corresponding graph state of the steady state of the hybrid Clifford circuit. We expect that in this free fermion system, there exists an Anderson localization transition with respect to the node degrees. When $p>p_c$, in the area law phase, the random graph has a local structure and fermions can only hop to the sites which are spatially nearby. Since we are considering the one dimensional random system, the single particle wave function is always spatially localized. In contrast, in the volume law phase $p<p_c$, the Hamiltonian has long range hopping terms and the wave function can become delocalized. 

To quantitatively characterize how extensive the wave functions are, we consider the IPR defined as
\begin{eqnarray}
I_q(L) = \sum_{i=1}^L |\Psi(x_i)|^{2q}.
\end{eqnarray}
where $\Psi(x_i)$ is a random eigenstate of $T_Z$. This quantity computes the $q$-th moment of the eigenstate coefficients and scales as $I_q \sim L^{-\tau_q}$ with $\tau_q = D_q (q-1)$. As we mentioned in the introduction, $D_q$ is the fractional dimension and has been used to distinguish between the extensive and localized states in the Anderson localization transition. In our model, when $p>p_c$, the system is localized and we have $D_q=0$ for all $q>0$. We are going to study $D_q$ for the volume law phase with $p<p_c$.

To numerically compute $\tau(q)$ and $D_q$ in the random system at finite system size, we need to take an ensemble average over $\Psi (x_i)$. The correct way to do this is to consider the quenched average over $\log I_q(L)$, i.e.,
\begin{align}
    \tau_q=-\frac{\langle \log I_q(L)\rangle}{\log L}.
    \label{eq:tau}
\end{align}
The average $\langle \cdot \rangle$ is taken over both different eigenvectors, $\Psi_k$, in one realization of $T_Z$ and different instances of the random Clifford circuit. 
We also compute the annealed average defined as
\begin{align}
    \tau_q^*=-\frac{\log \langle  I_q(L)\rangle}{\log L}. \label{eq:tau_star}
\end{align}
In random systems, this quantity is much easier to obtain analytically. Previous experience in random matrices and spin glass systems tell us that when the system is ergodic, the quenched average and annealed average give the same result \cite{Evers_RMP,Derrida_REM,Chamon}. On the other hand, in many non-ergodic systems, $\tau_q$ and $\tau^*_q$ can be quite different. 

By studying $I_q(L)$, we find that at any finite measurement rate $0<p<p_c$, the behavior of the graph states generated in the hybrid random Clifford circuits is dramatically different from the $p=0$ limit. To see this we numerically measure both $\tau_q(L)$ and $\tau^*_q(L)$ at general $q$ using the ensemble of graphs generated by the random Clifford circuits with up to $L=4000$ sites. We furthermore use these scaling exponents to extract the value of the fractal dimension $D_q$ in the thermodynamic limit $L \rightarrow \infty$.  In both cases, we will see that the usual volume law Clifford wave function in the absence of measurements ($p=0$) behaves as a fully extended wave function, while for any finite measurement rate there is evidence of multifractal scaling behavior.

\begin{figure}
    \centering
     \includegraphics[scale=0.50]{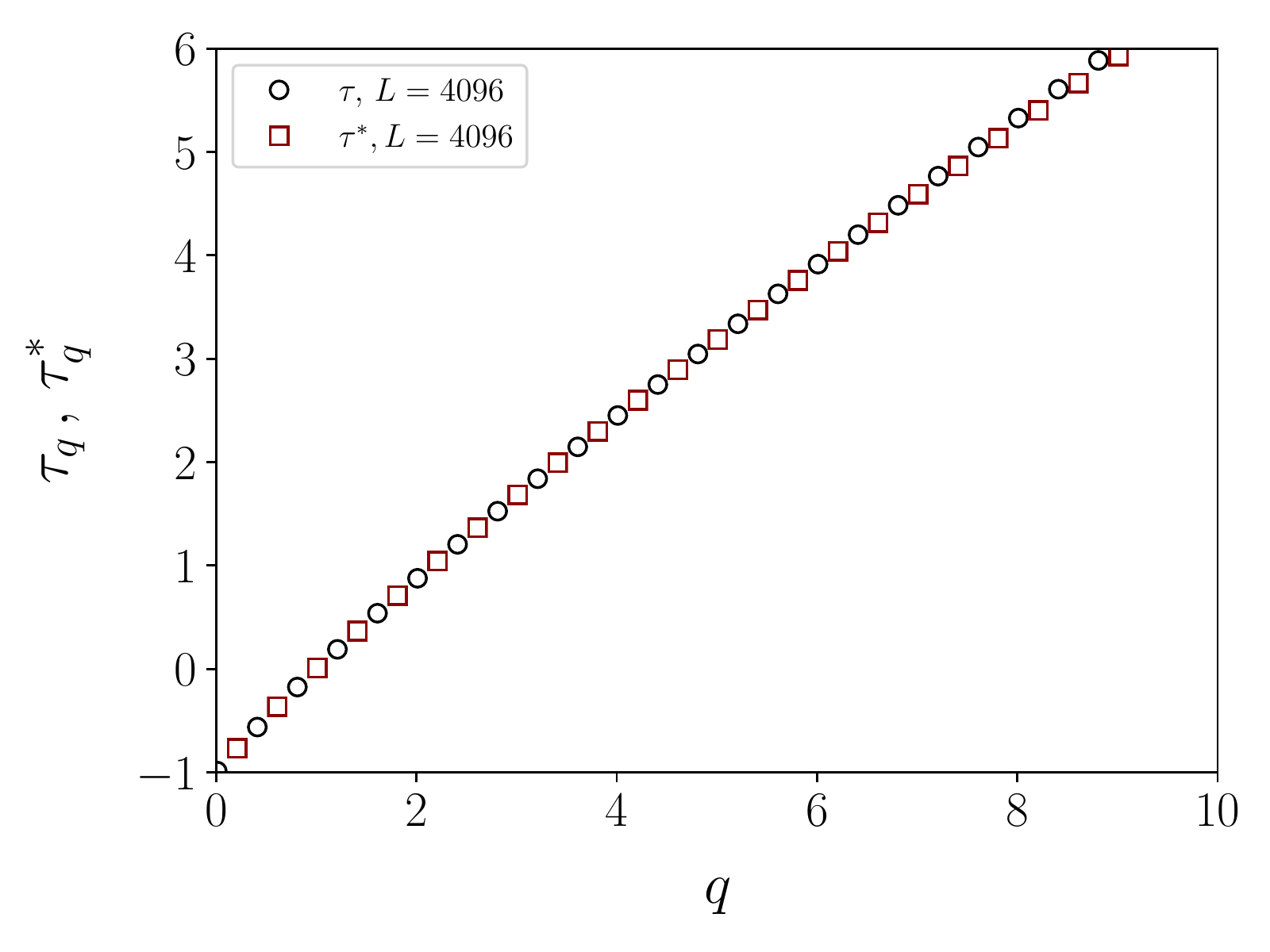}
    \includegraphics[scale=0.50]{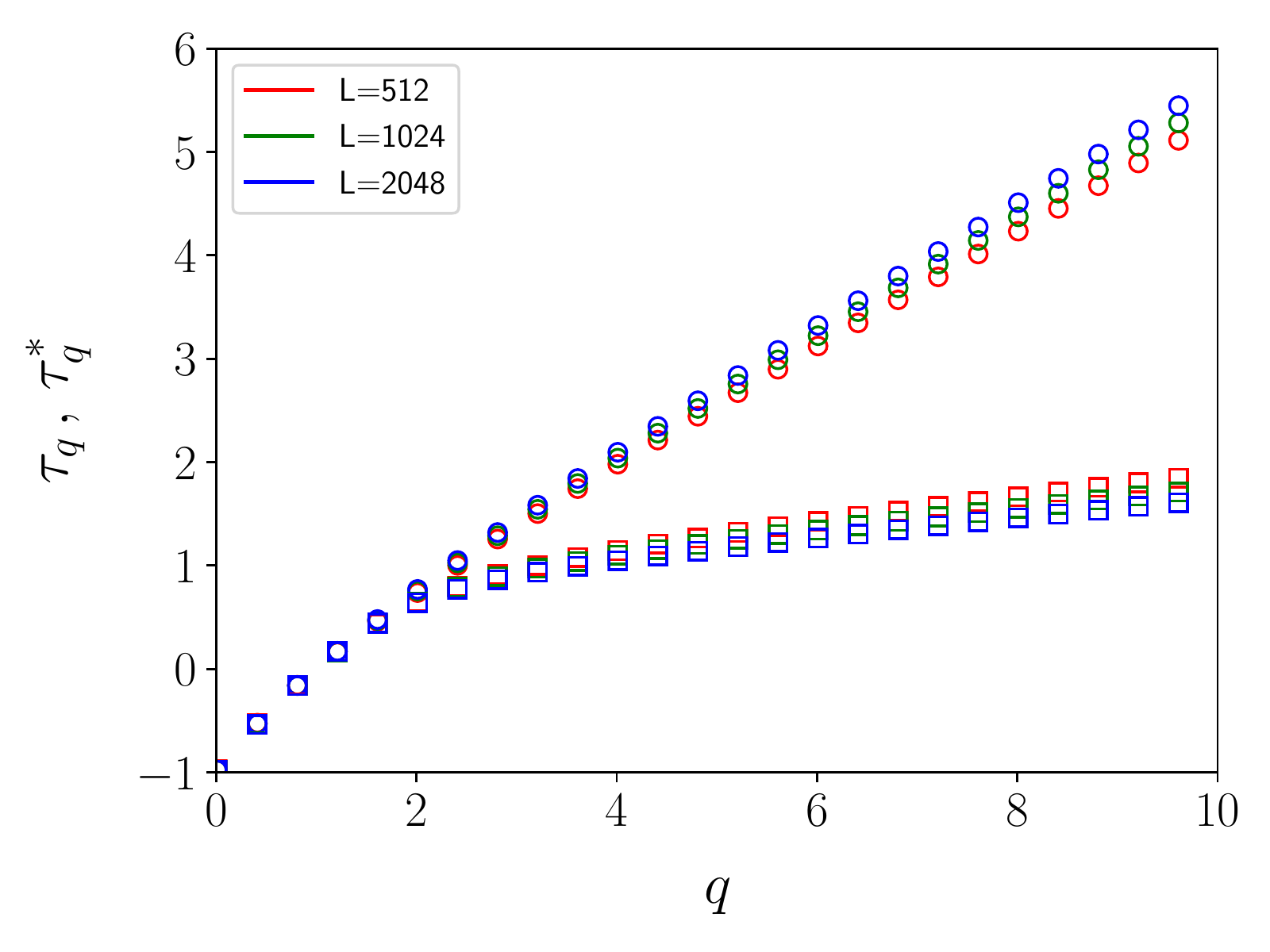}
    \caption{Comparison of the multifractal exponents $\tau_q$ (circles) and $\tau^*_q$ (squares) of the IPR as a function of $q$. (a) In the special case when $p=0$, both $\tau_q$ and $\tau^*_q$ are the same regardless of the value of $q$. (b) For non-zero measurement rate $p=0.05\ll p_c$, $\tau_q$ and $\tau^*_q$ show drastically different behavior when $q>2$. Their difference increases slightly as we increase the system size. For other values of $p<p_c$, similar behaviors are observed, providing evidence of the non-ergodic nature of the full volume law phase.}
    \label{fig:tau_q_star}
\end{figure}

\begin{figure}
    \centering
     \includegraphics[scale=0.5]{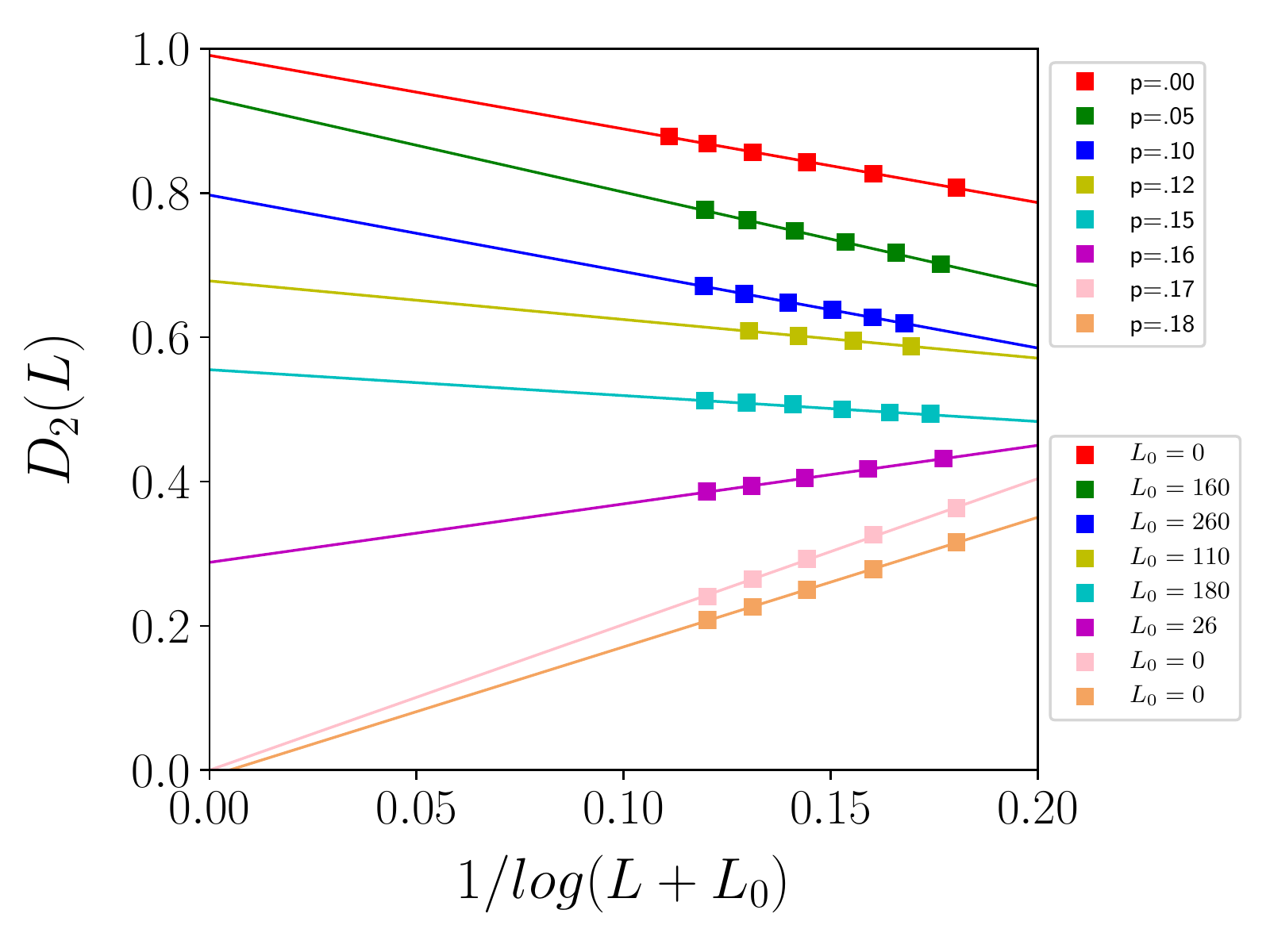}
    \includegraphics[scale=0.5]{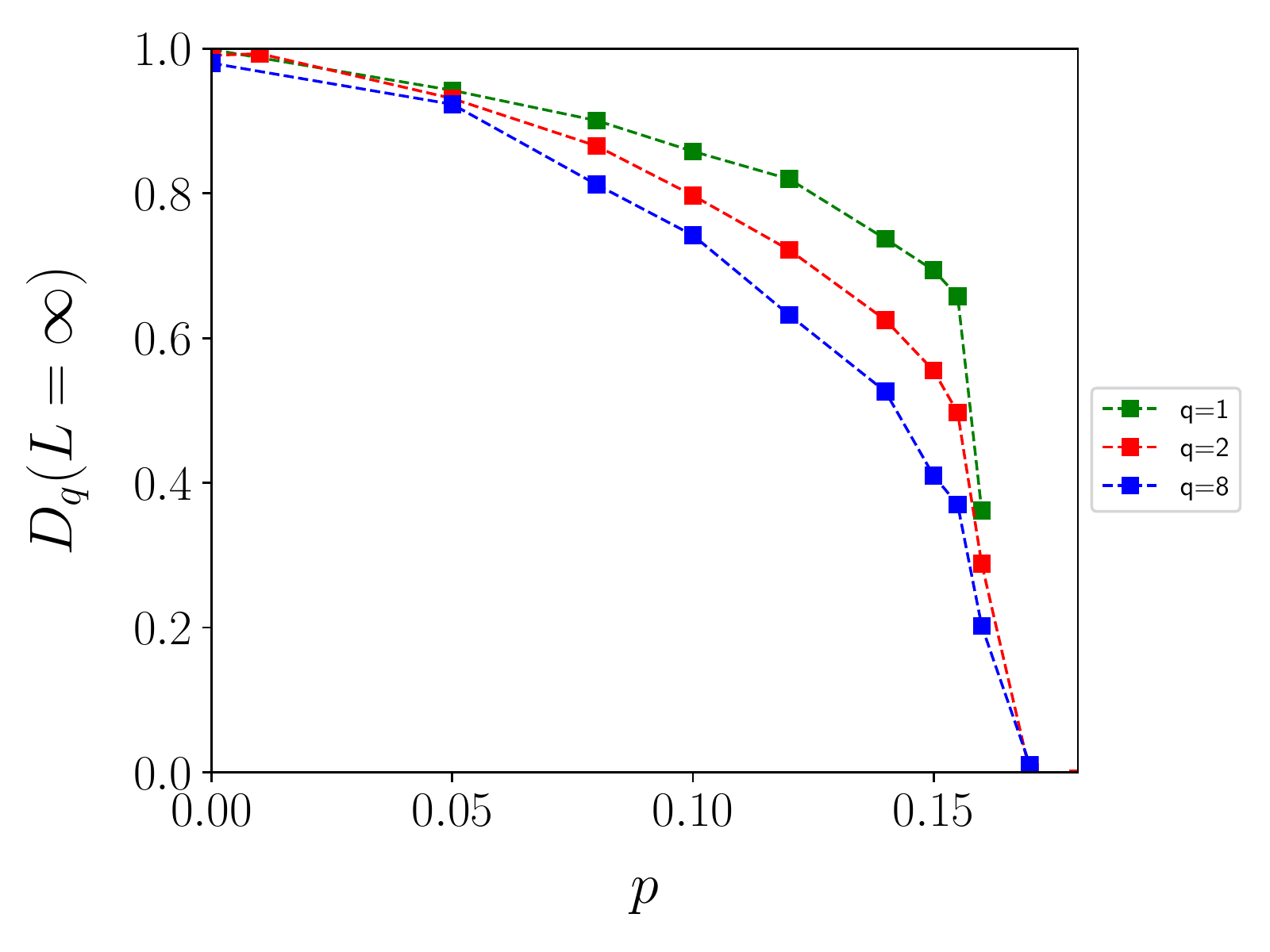}
    \caption{{(\it Top)} Finite size scaling of the fractal dimension $D_q = \tau_q/(q-1)$. The data appears to follow the form $D_q(L) \sim \frac{c}{\log(L+L_0)}+D_q(\infty)$. {\it (Bottom)} The fractal dimension extrapolated to the thermodynamic limit as a function of $p$ for different values of $q$. Note that the non-linearity as a function of $q$ is present throughout the volume law phase, providing evidence of a full {\it multi-fractal phase}. }
    \label{fig:Dq_p}
\end{figure}

\begin{figure}
    \centering
   \includegraphics[scale=0.45]{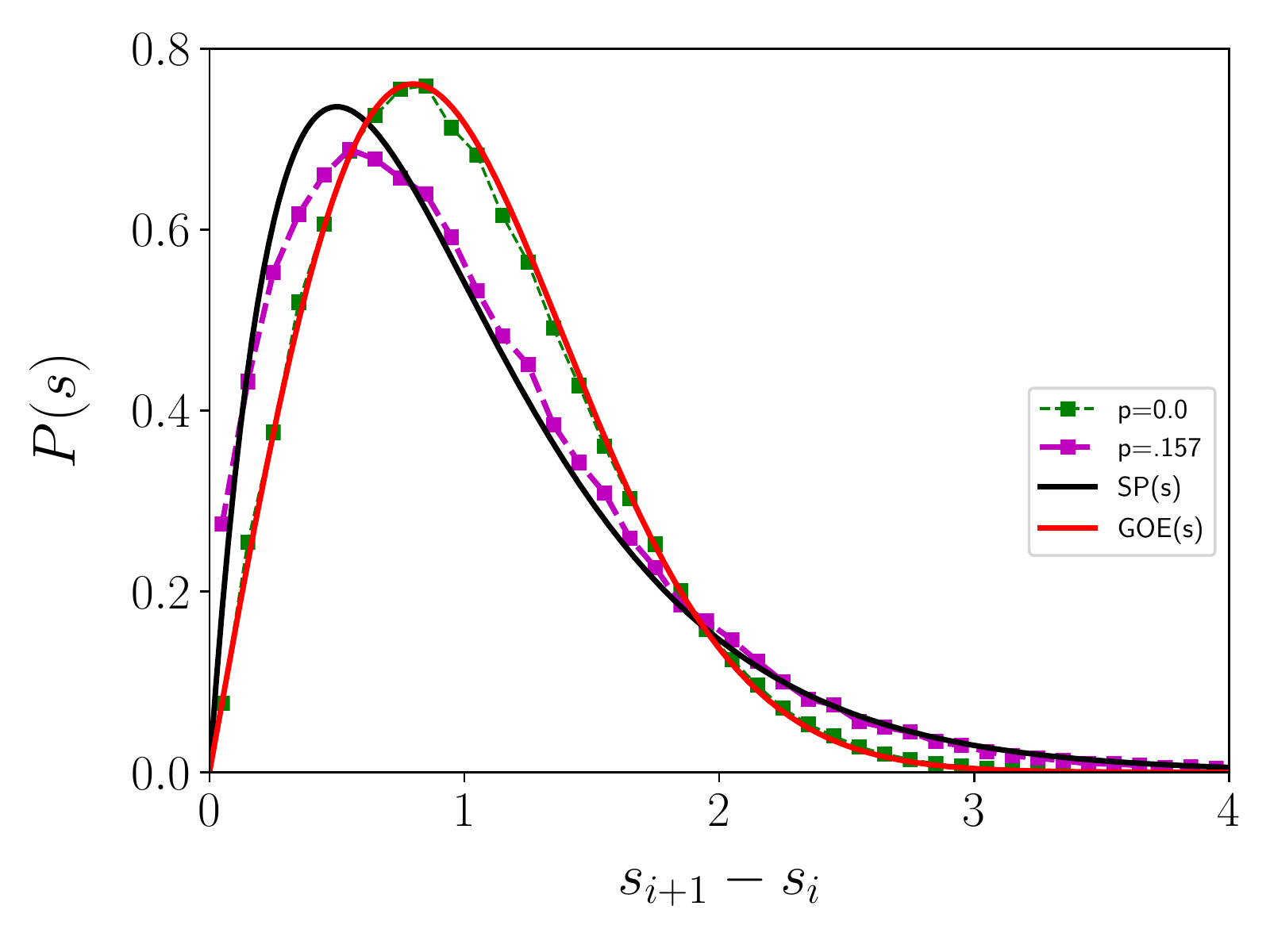}
    \includegraphics[scale=0.45]{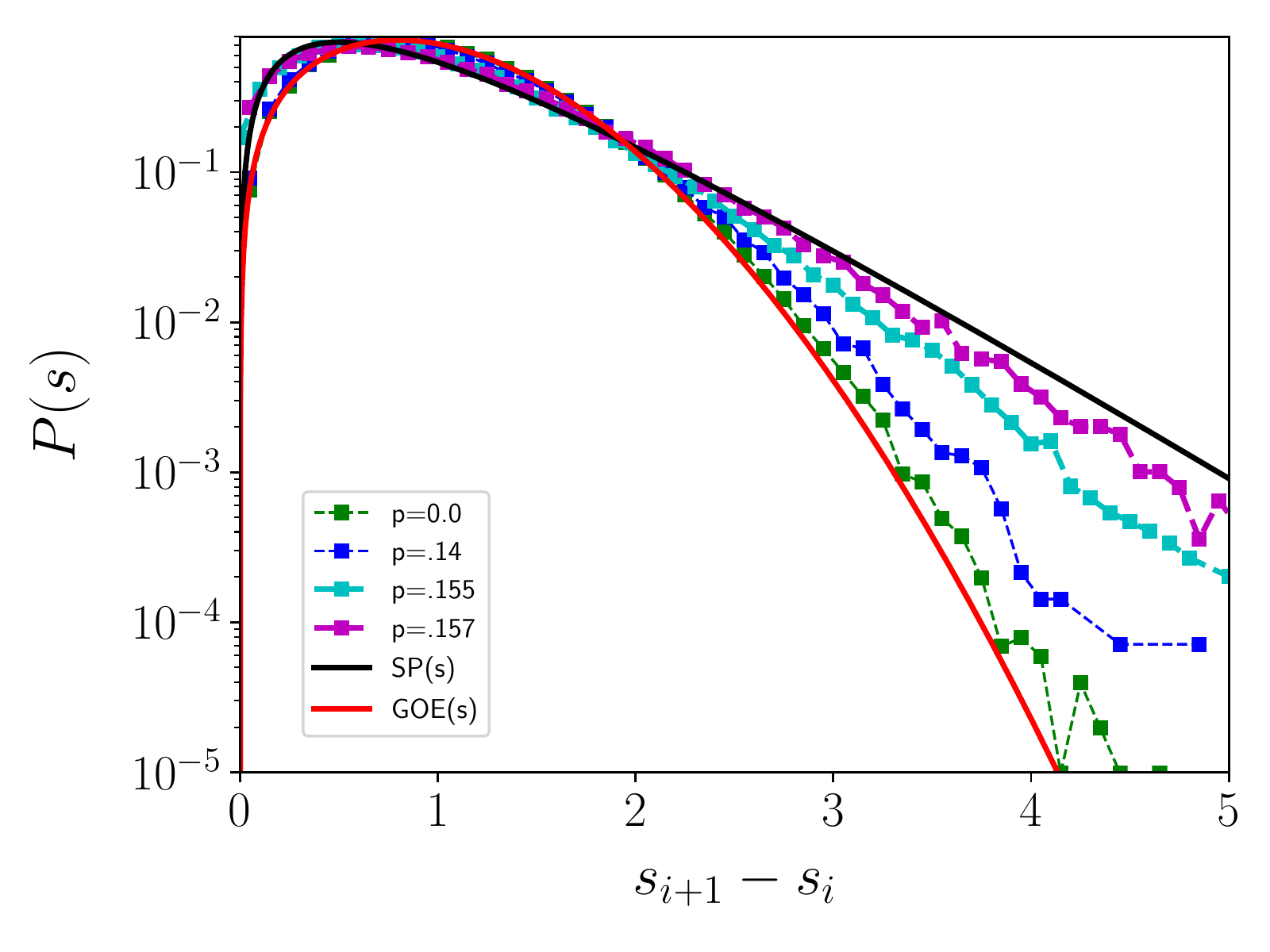}
    \caption{The eigenvalue level-spacing probability distribution of the adjacency matrices which result from the hybrid random Clifford dynamics. The unfolding procedure is applied only to eigenvalues in the range $\lambda_i \in [-20,-3]$, where the density of states is relatively smooth. $P(s)$ appears close to the GOE distribution deep in the volume law phase, but appears to approach the semi-Poisson distribution near the critical point. The top and bottom plots are the same results but on a linear scale and log scale respectively.}
    \label{fig:cliff_spacing}
\end{figure}

We first consider the limit $p=0$. We measure both $\tau_q$ and $\tau^*_q$, using Eq.'s \ref{eq:tau} and \ref{eq:tau_star}. As shown in Fig.~\ref{fig:tau_q_star}, we find that $\tau_q$ and $\tau^*_q$ take the same value for arbitrary $q$ (on top of each other in the plot). This indicates that when the measurement rate $p=0$, the single particle wave function is ergodic and $I_q(L)$ is a self-averaging quantity. We can further use $\tau_q$ to extract the fractal dimension $D_q$. We find that the finite size scaling of the fractal dimension very closely follows the form

\begin{align}
    D_q(L)\sim 1-\frac{f(q)}{\log L}.
    \label{eq:log_scaling}
\end{align}
The fractal dimension $D_q$ approaches 1 for all $q$ (See Fig.~\ref{fig:Dq_p} for the plot of $D_2(L)$ at finite $L$). This is again consistent with a single particle wave function which is fully ergodic and extensive in the thermodynamic limit. Note that the same finite size sub-leading correction of $D_q$ at finite $L$ is also observed in free fermion models which use the Gaussian orthogonal ensemble (GOE) random matrices as the Hamiltonians \cite{B_cker_2019}. Finally, we also examine the level spacing statistics of the eigenvalues in $T_Z$ by computing the probability distribution $P(s_i)$ of $s_i = e_i - e_{i+1}$, the spacing between adjacent eigenvalues. It is known that in GOE random matrices, $P(s)$ takes the following form
\begin{eqnarray}
P(s) = \frac{\pi}{2} s e^{-\frac{\pi}{4} s^2}.
\label{eq:goe}
\end{eqnarray} 
In Fig.~\ref{fig:cliff_spacing}, we plot the level spacing distribution for the ensemble of adjacency matrices with $p=0$.  To numerically obtain $P(s)$, we perform the unfolding procedure \cite{RevModPhys.53.385,PhysRevB.66.052416}, which compensates for the non-constant density of states in the eigenvalue distribution. We find that $P(s)$ very closely follows the same GOE form. 
Overall, we find that graphs states generated from random Clifford circuits without measurements are very well behaved. The associated adjacency matrix shares similar properties with a GOE random matrix and its eigenstates show behavior consistent with fully extended ergodic wave functions.

\begin{figure}
    \centering
     \includegraphics[scale=0.45]{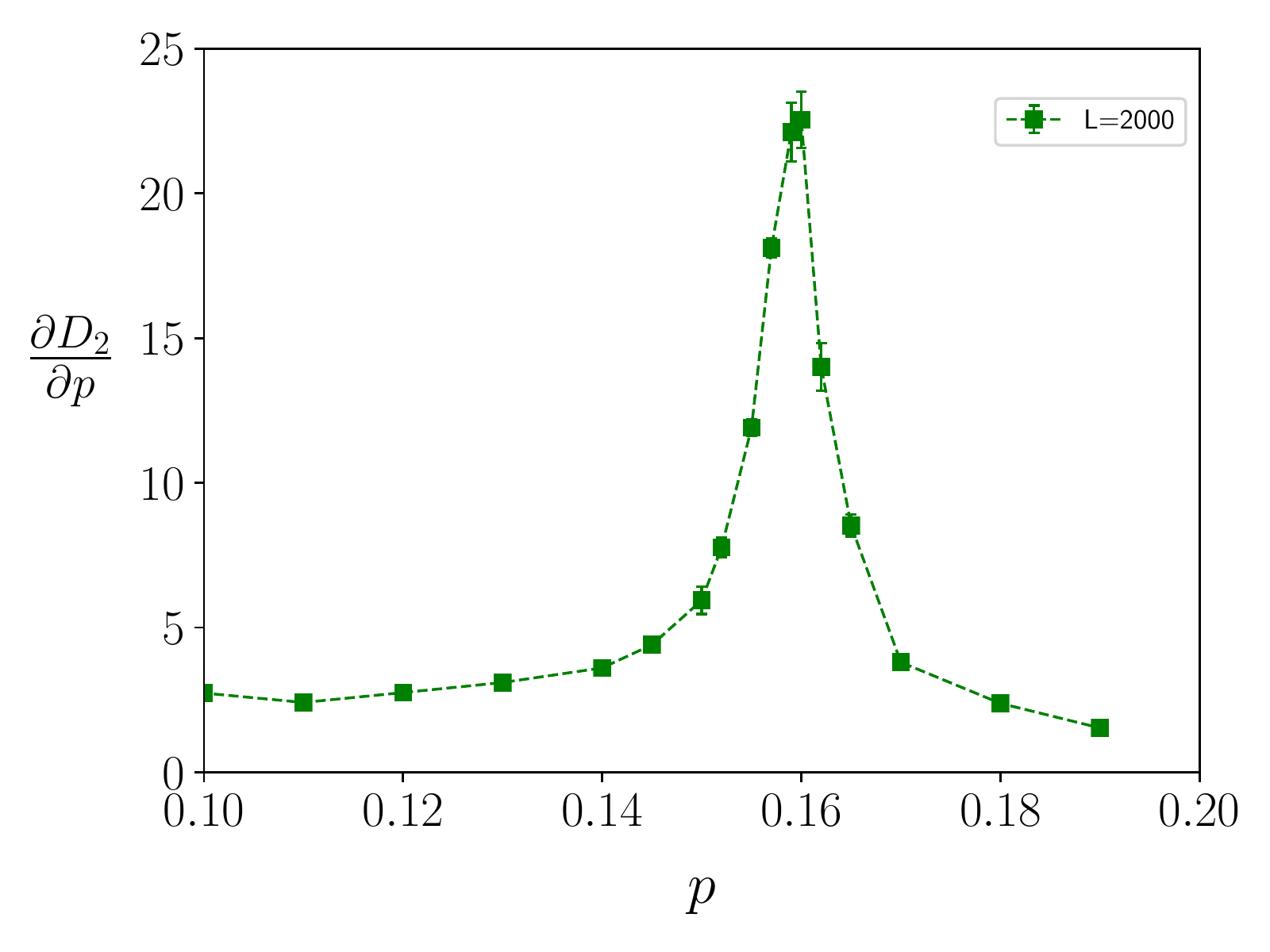}
    \caption{The derivative of the fractal dimension $D_2$ as a function of measurement rate $p$, for a fixed system size $L=2000$. We see that there is a peak in the derivative $\partial D_2(p) / \partial p$, at exactly the transition point $p_c\approx 0.16$.}
    \label{fig:derivative}
\end{figure}

We now consider the case of a finite measurement rate. The behavior of the wave functions for any $p>0$ is significantly different from the $p=0$ case. The quenched exponents $\tau_q$ behave very differently than the annealed exponents $\tau^*_q$ and the fractional dimension $D_q$ shows complicated multifractal behaviour. First consider the behavior of $\tau_q$ vs that of $\tau^*_q$, for some nonzero value of $p$ deep in the volume law phase. As shown in Fig.~\ref{fig:tau_q_star} (b), for $p=0.05$, $\tau_q$ and $\tau^*_q$ are equal only when the moment $q   \lesssim q_c=2$. For $q \gtrsim 2$, $\tau_q$ grows linearly with $q$ and there is a large gap between $\tau_q$ and $\tau^*_q$. Furthermore the gap $|\tau_q-\tau^*_q|$ grows with system size, indicating that this effect will persist in the thermodynamic limit.  This discrepancy between the quenched and annealed average exists for all $0 < p < p_c$ and indicates that the steady state for the non-thermal volume law phase is non-ergodic in the graph space.

We now focus on the behavior of the fractal dimension $D_q$ as a function of both $q$ and measurement rate $p$, within this non-thermal phase. For any finite size system, we measure $\tau_q$ using Eq.~\ref{eq:tau}, and determine the fractional dimension using the relationship $D_q = \tau_q /(q-1)$. Note that for $q=2$, $D_q = \tau_q$. We find that the fractal dimension at fixed $q$  decreases monotonically as we increase $p$. In particular,  as we cross the phase transition point, we observe a sharp peak exactly at $p_c\approx 0.16$ in the derivative of $D_q(p)$ (See Fig.~\ref{fig:derivative}) for $\partial D_2(p)/\partial p$ with fixed system size.


In order to calculate the precise value of the fractal dimension $D_q$ in the thermodynamic limit we must perform finite size scaling. The finite size results for $q=2$ are shown in Fig.~\ref{fig:Dq_p}.  We again find that there are logarithmic corrections in the finite size limit, and so a careful extrapolation to the thermodynamic limit must be performed. In fact, we find that the finite size effects are more significant for nonzero measurement rates. We empirically find a very good fit to the form
\begin{eqnarray}
D_q(L) = D_q(\infty) + \frac{f(q)}{\log(L+L_0)},
\label{eq:log_scaling_2}
\end{eqnarray}
where we include the additional fitting parameter $L_0$.  The data follows this scaling form for all systems sizes we measured from $L=256$ to $L\sim 4000$ sites. Note that $L_0$ is much smaller than the largest system size $L=4000$ in the fitting and as $L\rightarrow \infty$, we recover the same form as for the $p=0$ case. After extrapolating to the $L=\infty$ limit, we notice that in the range $0 < p < p_c$, $D_2(p)$ has a non-integer fractal value.  

Using this extrapolation method, we plot $D_q$ as a function of p in the thermodynamic limit. 
We show the results in Fig.~\ref{fig:Dq_p}, for $q=1,2$ and $8$. Note that in the $q=1$ case, $D_1(q)$ is defined by the limit of the equation $D_q = \tau_q/(q-1)$, where $\tau_1(p)=0$ due to the normalization of the wave fucntion. We find that for all $q$, $D_q(p)$ forms a continuous curve which interpolates between $D_q=1$ at $p=0$ and $D_q=0$ at $p=p_c$.  Importantly, at all finite measurement rates in the volume law phase, the fractal dimension exhibits a strong $q$ dependence. 
That is, for any nonzero measurement rate in the volume law phase, the resultant graph states show multifractal behavior which is not present in the usual volume law phase without measurement. This multifractal behavior is reminiscent of the critical behavior of wave functions near an Anderson localization transition.

\begin{figure}
    \centering
     \includegraphics[scale=0.45]{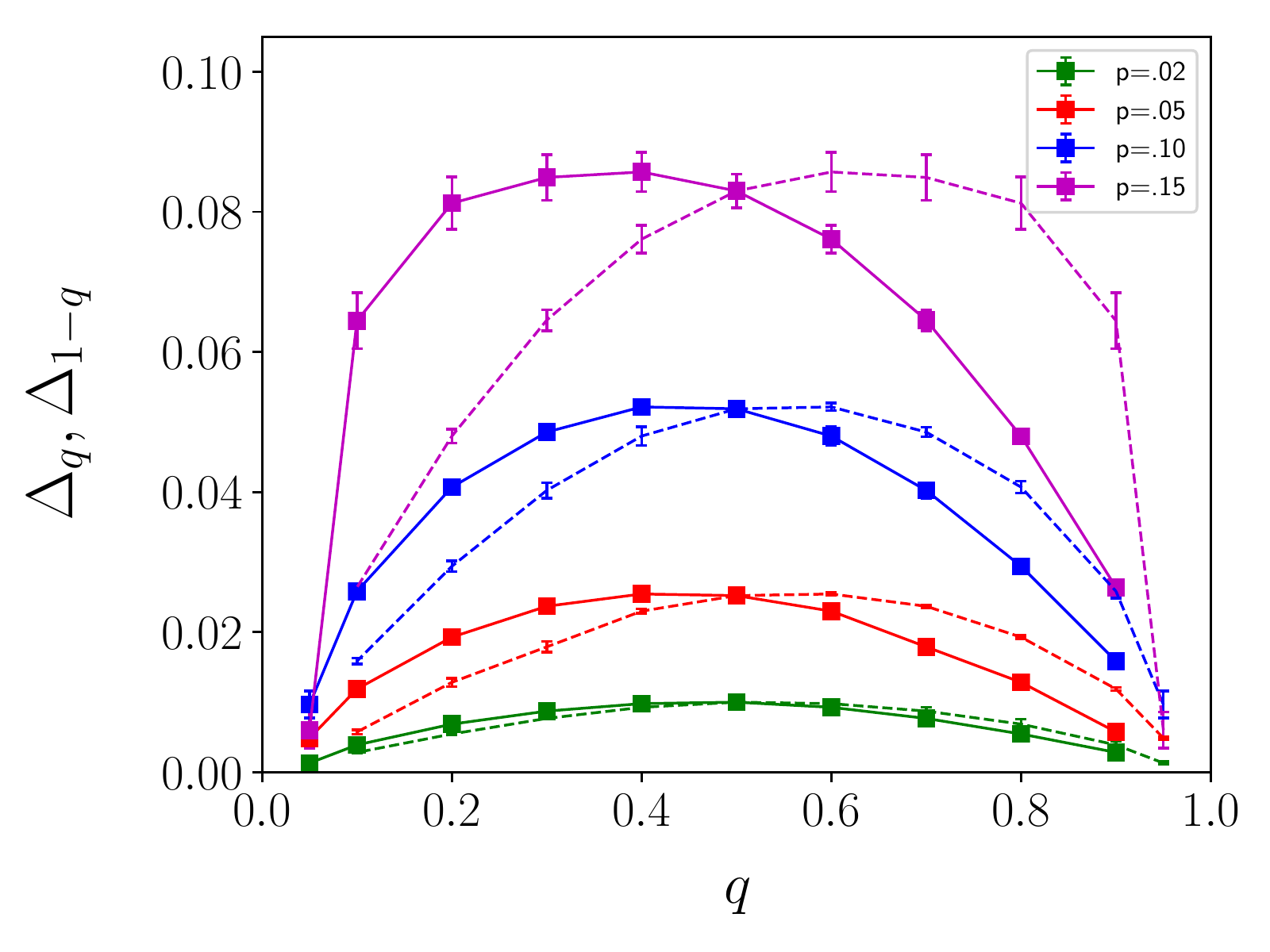}
    \includegraphics[scale=0.45]{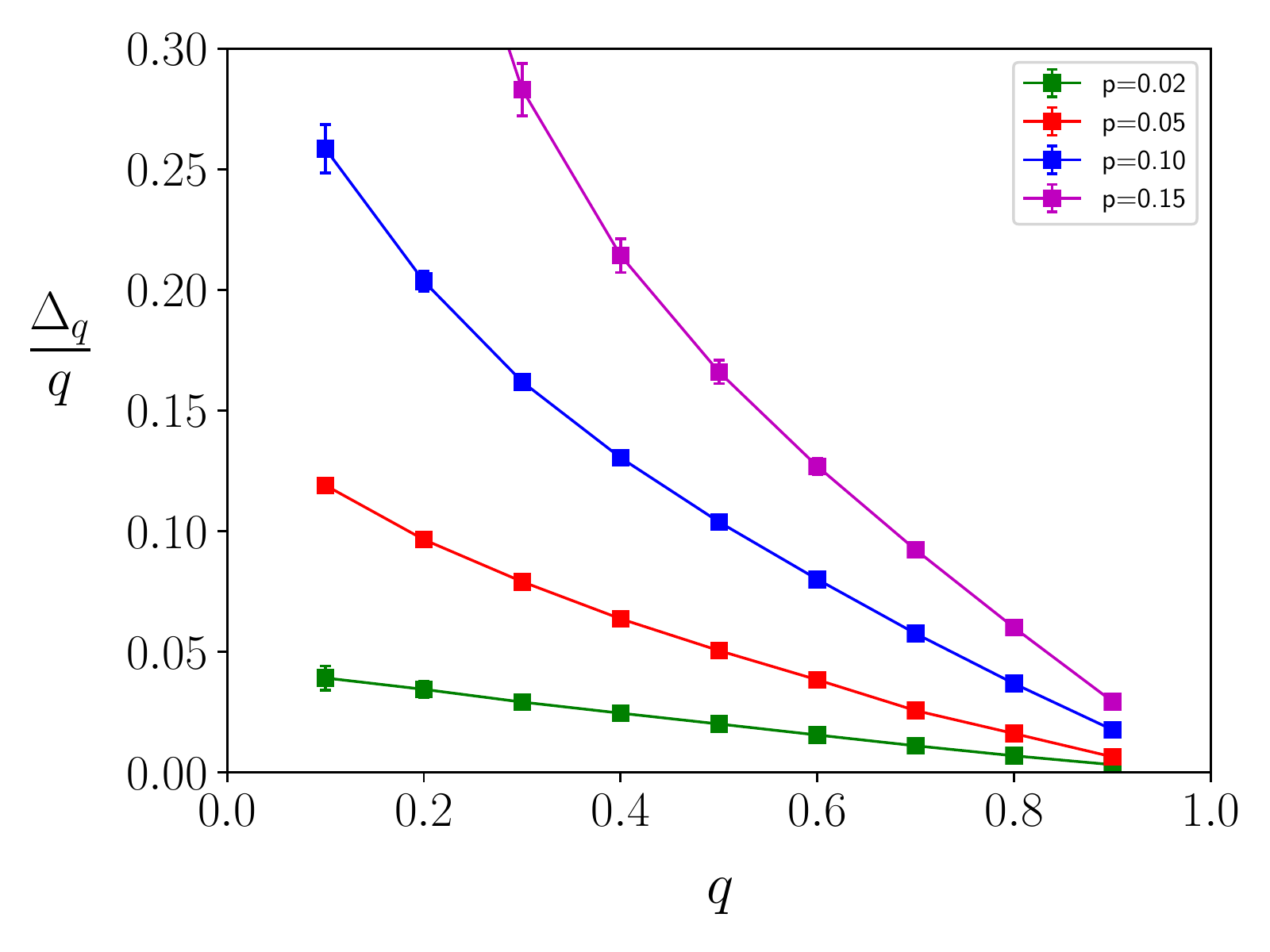}
    \caption{{(\it Top)} The anomalous dimension $\Delta_q = q-1-\tau_q$ (solid line) and $\Delta_{1-q}$ (dashed line) for small $q<1$, at different measurement rates $p$.  {\it (Bottom)} We plot $\Delta_q/q$ vs $q$. We find that for small $p$, this quantity is linear in $q$, ($\Delta_q/q \sim 1-q$) consistent with predictions for a weak multifractal system.}
    \label{fig:delta_cliff}
\end{figure}

We also look at the non-linearity of the multifractal exponent $\tau_q$ for $0<q<1.0$. Previously, the field theory calculation at the critical point of the Anderson localization transition suggests that the anomalous dimension $\Delta_q\equiv (q-1)-\tau_q$, which is defined as the deviation of $\tau_q$ from the fully ergodic case,  is symmetric around $q=1/2$ \cite{Mirlin_PRL,Evers_RMP,Gruzberg}. In Fig.~\ref{fig:delta_cliff} (a), we plot $\Delta_q$ and $\Delta_{1-q}$ as a function of $q$ in the thermodynamic limit and we find that they are close to each other. The difference between them is small for small $p$ and slightly increases as we increase $p$. When $p$ is close to zero, we find that $\Delta_q/q$ is a linear function of $q$, indicating that $\Delta_q \sim q(1-q)$. This parabolic form of $\Delta_q$ has also been observed in the critical wave function with weak disorder \cite{Chamon,Evers_RMP,Mirlin_PRL}.


Finally, we once again consider the eigenvalue spacing statistics of the adjacency matrix, $T_Z$, for nonzero measurement rate. For free-fermion models which undergo an Anderson localization transition, there is a qualitative change of the level spacing statistics as one moves across the critical point. In our model, as we mentioned previously, when $p=0$, the nearest neighbor level spacing distribution $P(s)$ is described by GOE. On the other hand, as shown in Fig.~\ref{fig:cliff_spacing}, near the critical point at $p=p_c$, the distribution $P(s)$ exhibits significant difference from GOE and is close to a semi-Poisson distribution \cite{Geraedts_2016,PhysRevE.59.R1315}, 
\begin{eqnarray}
P(s) = 4se^{-2s}.
\end{eqnarray}
Namely, there exists level repulsion between adjacent energy levels as $P(s\rightarrow 0) = 0$, and the tail of the distribution of $P(s)$ appears to decay exponentially as $\sim e^{-s}$. There appear to be small deviations from the exact semi-Poisson distribution in the intermediate regime.  We also compute $P(s)$ in the volume law phase with $0<p<p_c$. Deep inside the volume law phase, $P(s)$ appears to be described by the GOE distribution. For measurement rates closer to the critical point, the tail of the spacing distribution decays with some form between the GOE and semi-Poisson distribution. This deviation from GOE might be due to finite size effects and we expect that in the volume law phase, $P(s)$ becomes GOE in the thermodynamic limit.

\subsection{Hybrid random Clifford quantum automaton circuit}
\label{sec:automaton}

We now consider a hybrid circuit in which the unitary dynamics is composed solely of gates which preserve the computational basis. These are known as quantum automaton (QA) circuits, and have been studied in Ref's \onlinecite{iaconis2020measurement,iaconis2020quantum}. The general form of a QA gate is given by
\begin{eqnarray}
U_{QA} \ket{m} = e^{i\theta_m} \ket{\pi(m)},
\end{eqnarray}
where $\pi(m)$ is the permutation group on the $2^N$ basis states $\ket{m}$. When acting on an initial product state which has all spins perpendicular to the computational basis, $U_{QA}$ can create complex highly entangled wave functions \cite{iaconis2020measurement,iaconis2020quantum}. In particular, we have
\begin{eqnarray}
\ket{+} &=& \frac{1}{2^N} \sum_n \ket{n} \\
U_{QA} \ket{+} &=& \frac{1}{2^N} \sum_n e^{i\theta_n} \ket{n} .
\end{eqnarray}
When a finite rate of non-unitary composite measurements (explained below) are added to $U_{QA}$, there is again a phase transition between a volume law and and area law phase. In this case, the universality class of the critical point is distinct from that of the hybrid random Clifford circuit discussed in Sec.~\ref{sec:adjacency} which possesses an emergent conformal symmetry and therefore has critical exponent $z=1$ \cite{li2020conformal}. It was shown in Ref.~\onlinecite{iaconis2020measurement} that the critical point of the hybrid QA circuit falls exactly in the directed percolation universality class. This is a well known non-equilibrium critical point which has a dynamical critical exponent $z=1.581$. 

\begin{figure}
    \centering
     \includegraphics[scale=0.35]{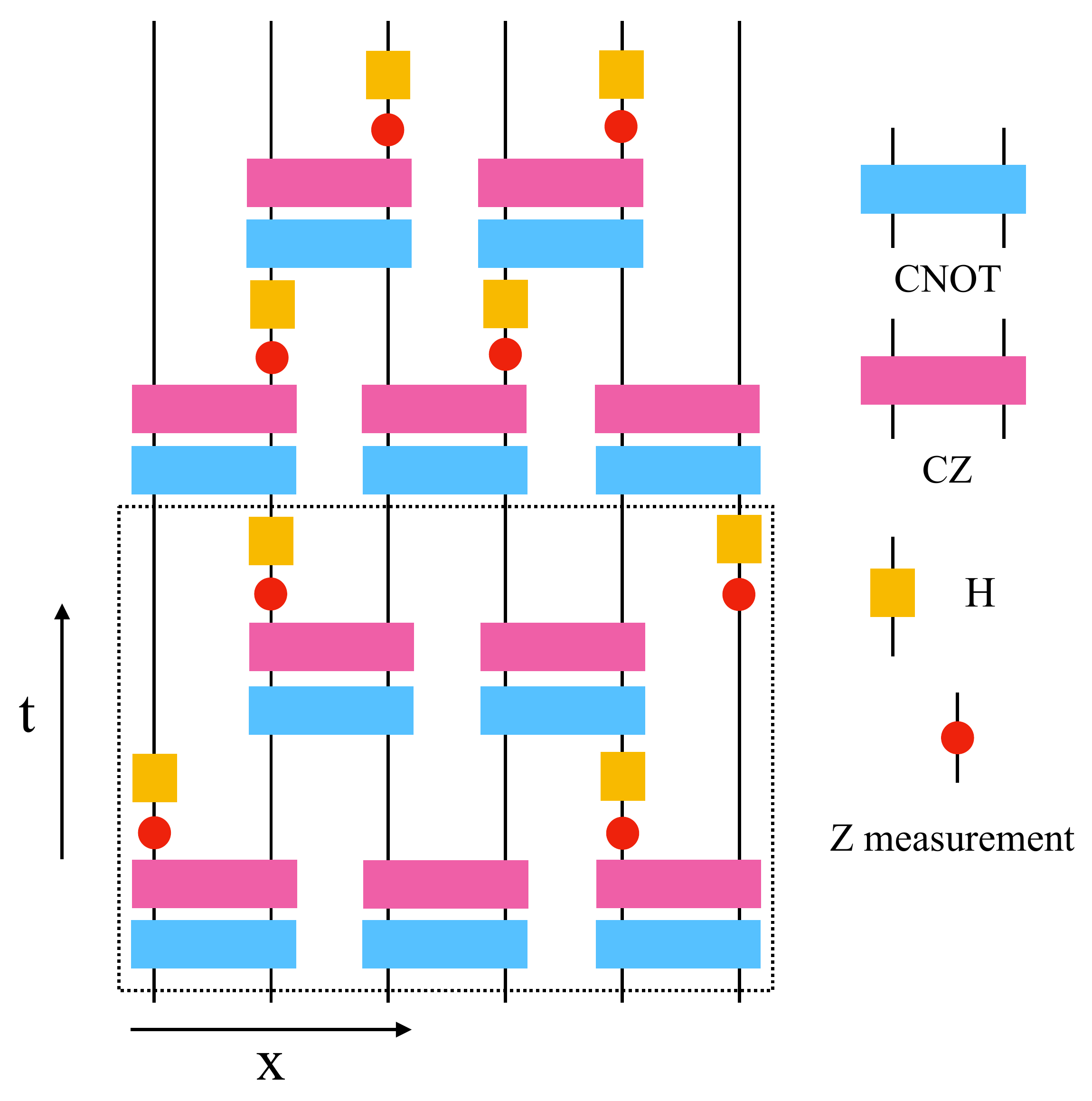}
    \caption{The cartoon for the Clifford QA circuit subject to Z measurement followed by Hadamard gate. The dashed box denotes the time evolution in one time step. The unitary QA circuit is composed of CNOT and CZ gates. There are two types of CNOT gates (control qubit on the left or the right qubit) and we apply them randomly with equal probability.}
    \label{fig:cartoon_QA}
\end{figure}

In what follows, we consider the case where the hybrid QA circuit contains unitary gates which belong to the Clifford group. We include a finite rate of composite measurements, whereby a spin is projectively measured in the Pauli Z basis followed by a Hadamard rotation (see Fig.~\ref{fig:cartoon_QA}). A Clifford QA circuit of this form was also studied in Ref.~\onlinecite{iaconis2020measurement}, and found to display the same characteristics as the generic QA hybrid circuit. In this model, there exists an entanglement phase transition between a volume law and area law phase at $p_c=0.075$. We now study the multifractal properties of the adjacency matrices formed by the steady state graph states of the hybrid Clifford QA circuits. Notice that in this model, the wave function is a graph state up to single qubit unitary S rotations (H rotation is not required).

\begin{figure}
    \centering
     \includegraphics[scale=0.45]{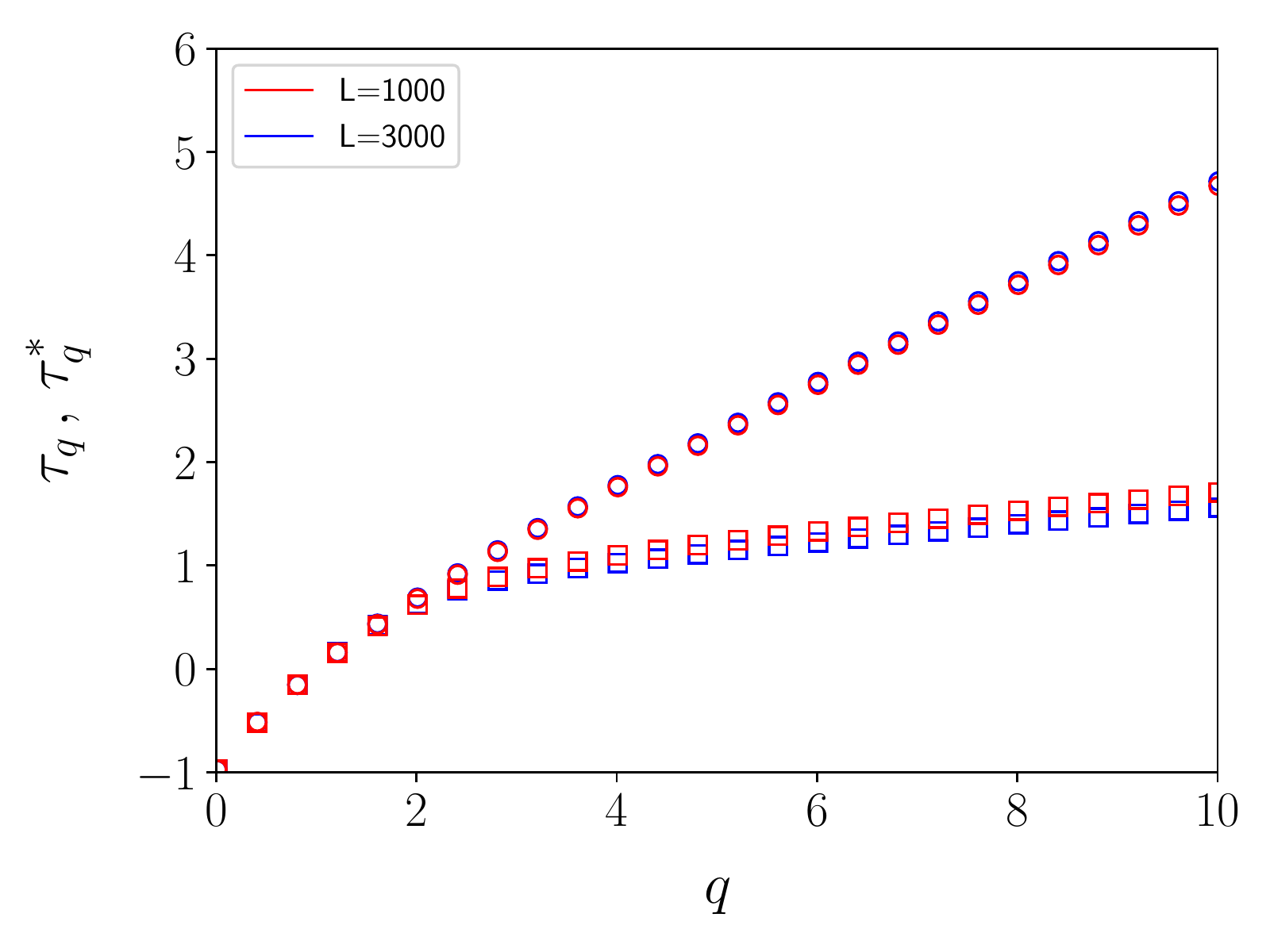}
    \caption{The IPR scaling exponent $\tau_q$ and the annealed scaling exponent $\tau^*_q$ for the QA circuit at $p=0.06$ in the non-thermal volume law phase. Once again, we find that for any finite measurement rate, $\tau$ and $\tau^*$ show dramatically different behavior at large $q$. }
    \label{fig:QA_tau}
\end{figure}

\begin{figure}
    \centering
    \includegraphics[scale=0.45]{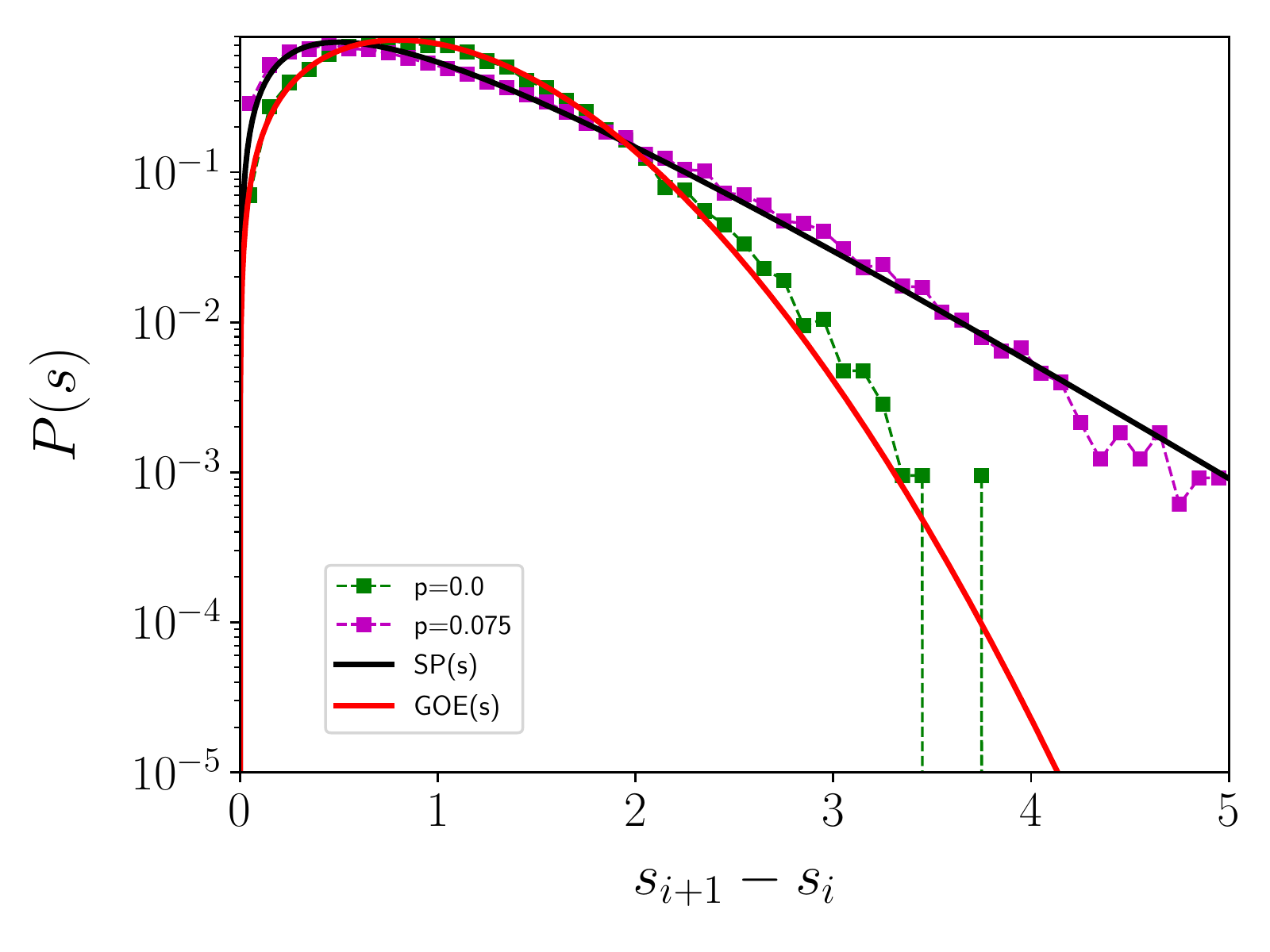}
    \includegraphics[scale=0.45]{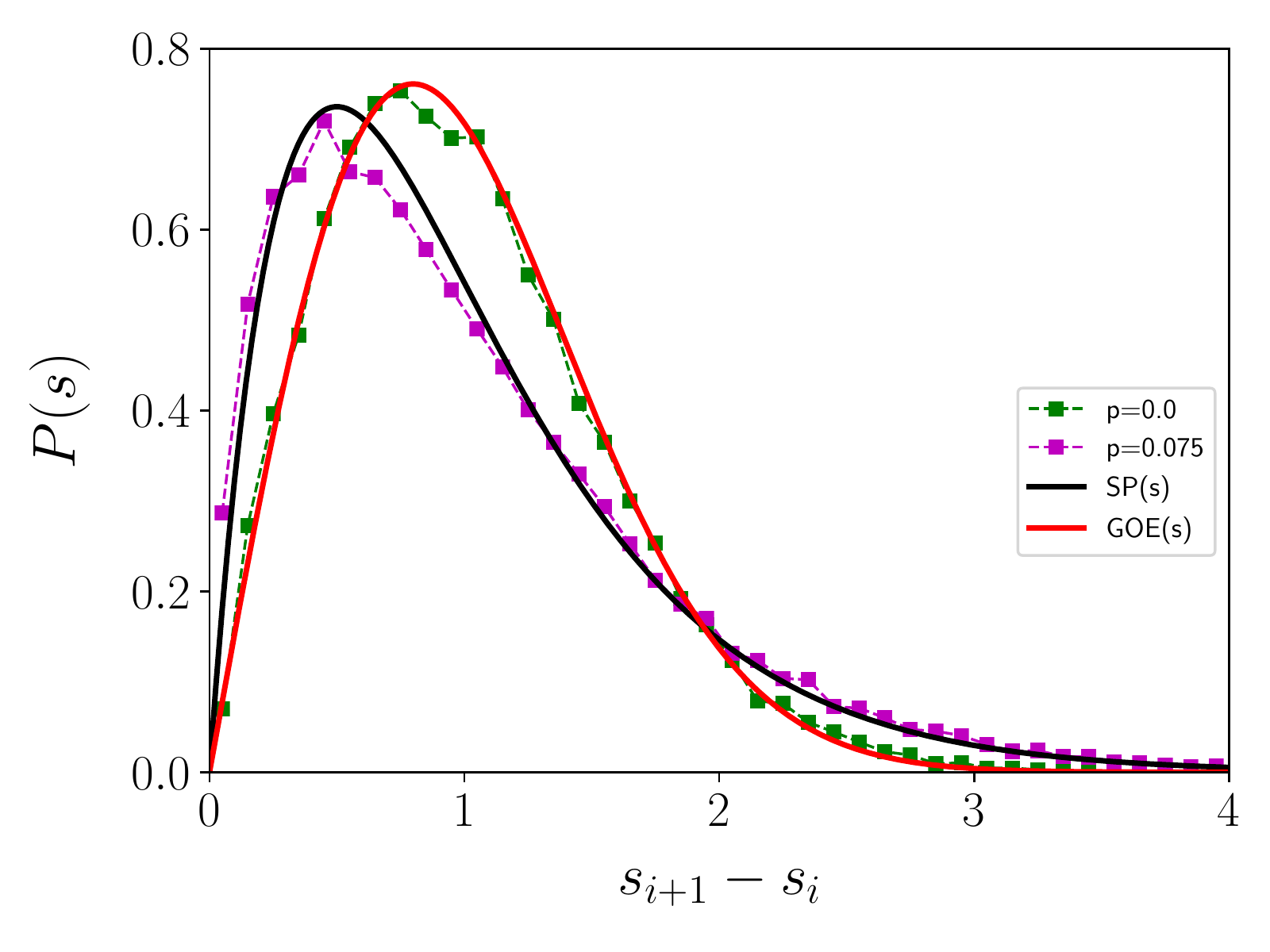}
    \caption{The unfolded level spacing statistics of the adjacency matrix for the QA Clifford wave functions at $p=0.0$ and at the critical point $p=0.075$. We again find that the level spacing distribution of the adjacency matrix appears to follow a GOE distribution for $p=0$ and is close to the semi-Poisson distribution at the critical point. }
    \label{fig:spacing_QA}
\end{figure}

 When the measurement rate $p=0$, the resulting eigenstates of the adjacency matrix with respect to the node degrees are fully extended and ergodic. In particular the quenched and annealed multifractal exponents are the same, i.e., $\tau_q = \tau^*_q$.  Using the same scaling form as in Eq.~\ref{eq:log_scaling}, in the thermodynamic limit, we find that $\tau_q \approx q-1$ 
with $D_q = \tau_q/(q-1) \approx 1$ for all $q$.  In Fig.~\ref{fig:spacing_QA}, we plot the eigenvalue level spacing distribution. We find that for QA circuits without measurements, the distribution follows the GOE form given by Eq.~\ref{eq:goe} with the wave function being fully ergodic and extended.

On the other hand, for any finite measurement rate, we again find that all steady state wave functions in the volume law phase display multifractal behavior. In addition, at large $q$, there is a large discrepancy between the quenched exponent $\tau_q$, and the annealed exponent $\tau_q^*$. This can be seen in Fig.~\ref{fig:QA_tau}, where at $p=0.06$, there is a large gap between $\tau_q$ and $\tau^*_q$ at large $q$, and this gap increases as we move towards the thermodynamic limit. This signals that there is a breakdown of ergodicity also in this QA model.

\begin{figure}
    \centering
     \includegraphics[scale=0.45]{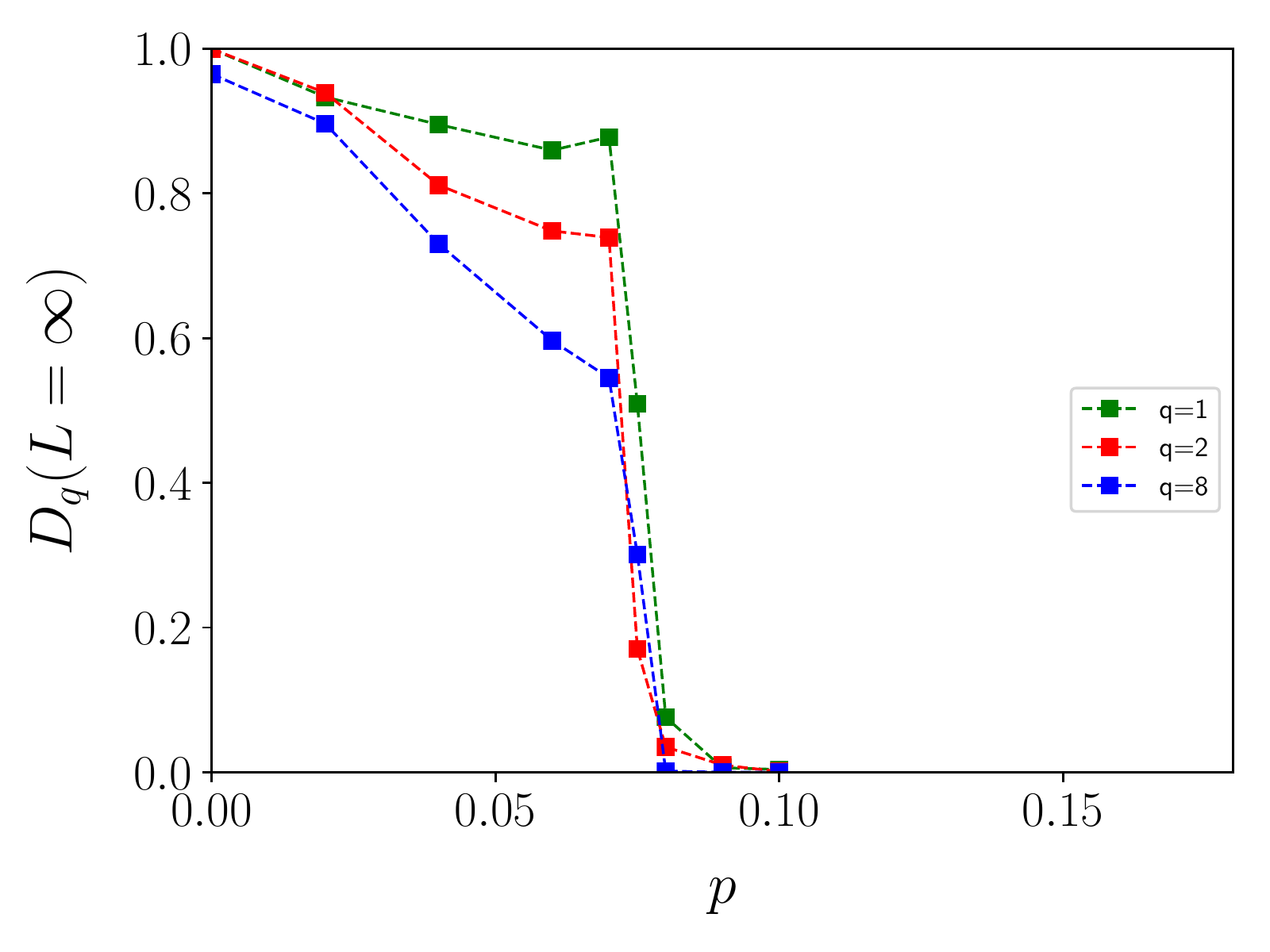}
    \caption{The fractal dimension $D_q$ for different measurement rates, $p$, in the thermodynamic limit, for the QA circuit. Note that the finite size effects appear to be larger in the QA circuit and performing the scaling to the thermodynamic limit is less precise than in the random Clifford circuit.}
    \label{fig:Dqp_QA}
\end{figure}

We further extract the fractal dimension $D_q$ as a function of measurement rate for different values of $q$. We use the scaling form in Eq.~\ref{eq:log_scaling_2}, to extrapolate $D_q$ to the thermodynamic limit. We show these results in Fig.~\ref{fig:Dqp_QA}, where we see very similar behavior as with the random Clifford circuit in Sec.~\ref{sec:adjacency}. In particular, $D_q$, when scaled to the thermodynamic limit, takes a non-integer fractal value for $0<p<p_c$. This fractal dimension goes to zero at the critical point and remains zero in the area law phase
for all $q$. Importantly, in the volume law phase, the value of $D_q$ has a strong $q$ dependence. This indicates that, again, the wave functions show complicated multifractal behavior throughout the entire volume law phase. Note that the finite size effects are somewhat larger in the QA model, and therefore the extracted curves for $D_q(L=\infty)$ are not as smooth as in the random Clifford model in Sec.~\ref{sec:adjacency}.

\begin{figure}
    \centering
     \includegraphics[scale=0.45]{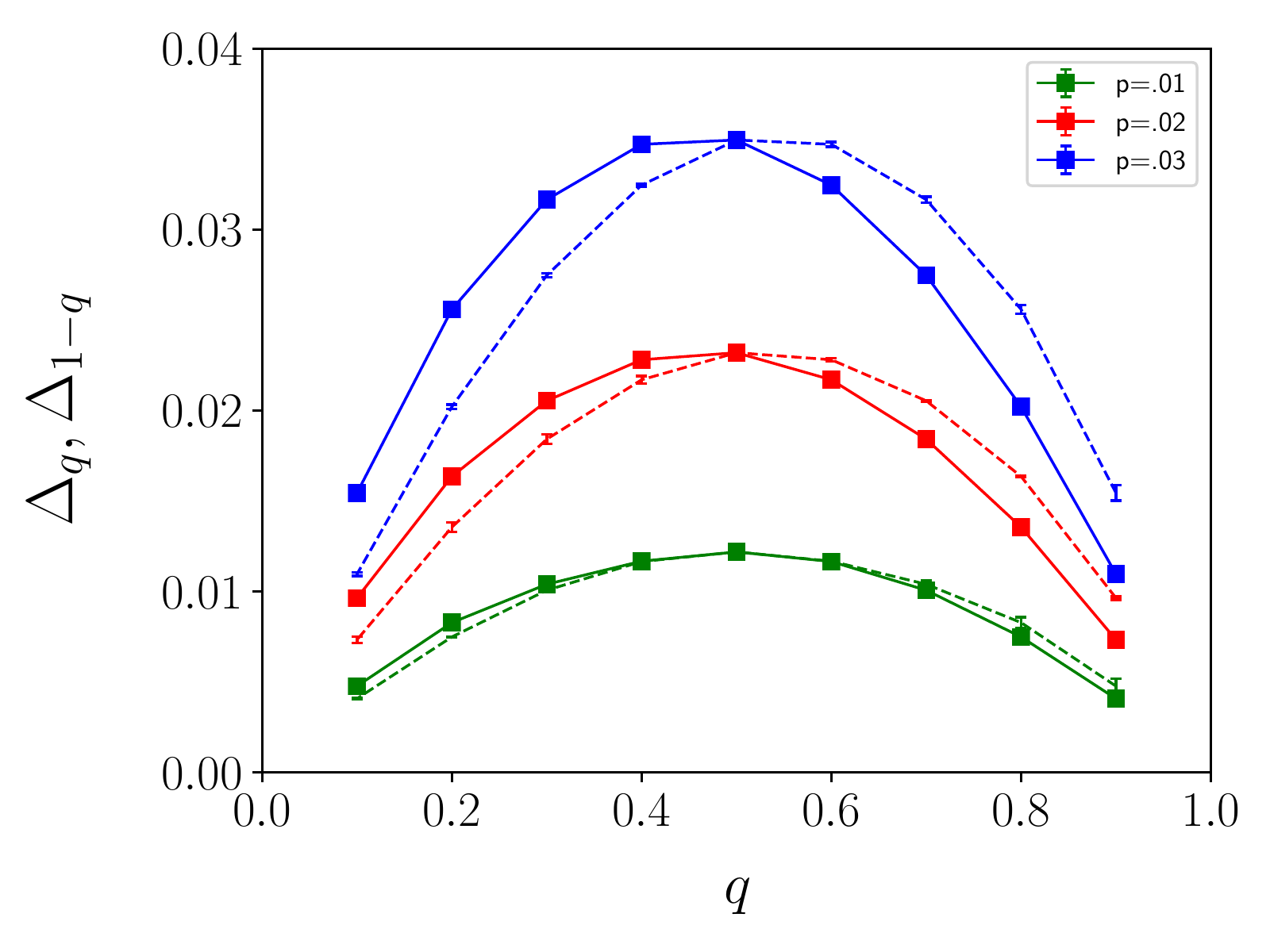}
      \includegraphics[scale=0.45]{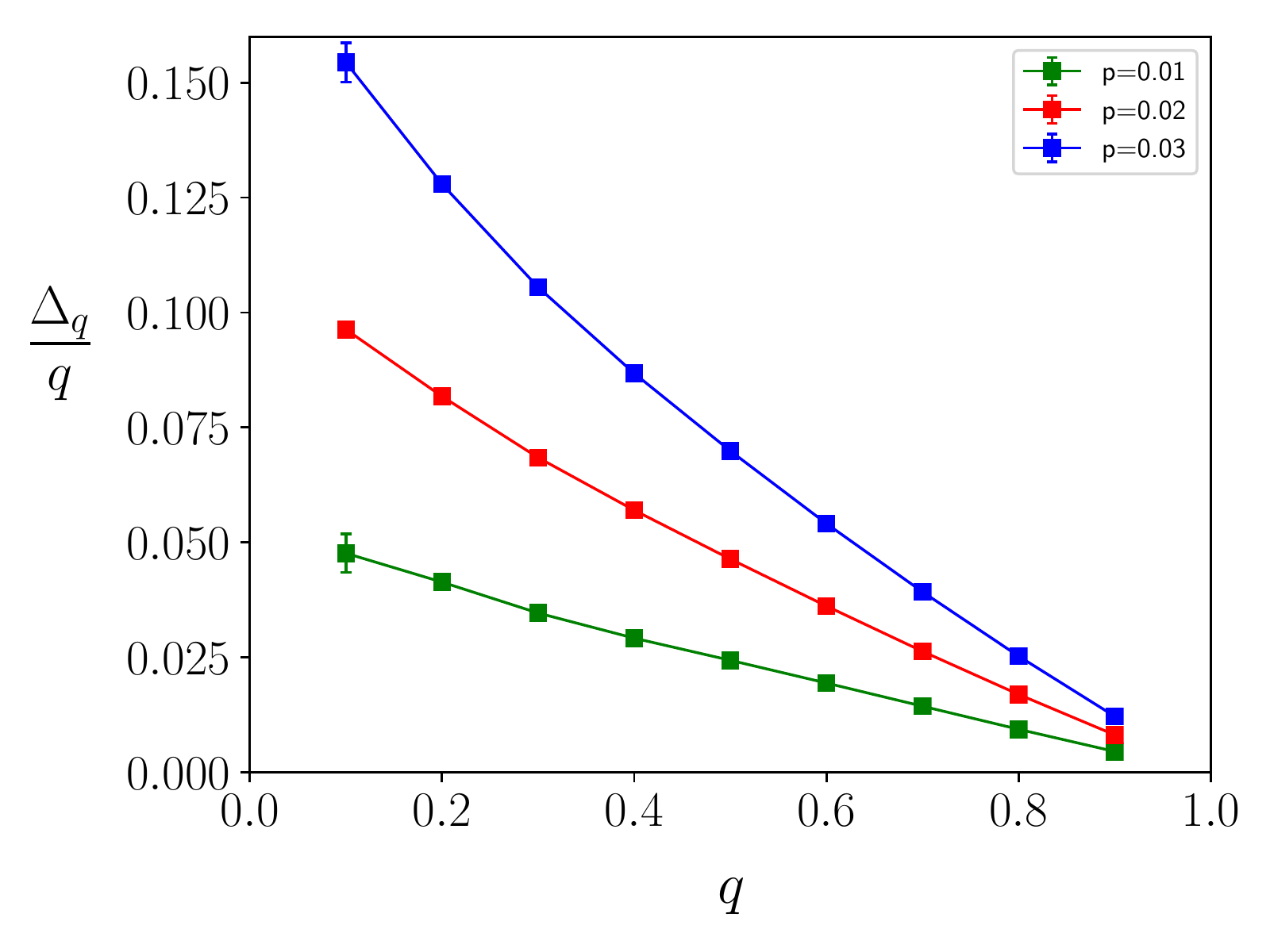}
    \caption{{\it (Top)} The anomalous dimension $\Delta_q$ and $\Delta_{1-q}$ for small $q<1$ in the QA circuit.  For measurement rate $p$ close to zero, $\Delta_q \approx \Delta_{1-q}$. {\it (Bottom)} We also show $\Delta_q/q$, and find that for small $p$, this is linear is $q$, which is again consistent with predictions for weak multifractal behavior.}
    \label{fig:delta_QA}
\end{figure}

In Fig.~\ref{fig:delta_QA}, we plot both the anomalous dimension $\Delta_q$, and $\Delta_q/q$ for $0<q<1$. For small measurement rates, $\Delta_q \approx \gamma q(1-q)$, for some small constant $\gamma$. This nonlinear functional form also follows the predictions for weak multifractal systems.

ummWe finally also study the level spacing statistics near the critical point $p_c$ in the Clifford QA model. The results are show in Fig.~\ref{fig:spacing_QA}.  We find that the ensemble of adjacency matrices show level spacing statistics which are very close the the semi-Poisson distribution.  The tail of the distribution decays like $e^{-s}$, and due to level repulsion $P(s)$ goes to zero as $s\rightarrow 0$. In summary, we also find multifractal behavior in the steady state wave functions of  the non-thermal volume law phase of the random Clifford QA circuit.  In  future works, the more general QA model might provide a tractable platform for extending these results to non-Clifford systems.

\section{non-unitary random free fermion dynamics}
\label{sec:fermion}
We consider the one dimensional discrete non-unitary free fermion circuit defined in Ref.~\onlinecite{Chen_2020}. The (unnormalized) non-unitary time evolution operator is defined as 
\begin{align}
    U=\prod_{t=1}^TU_\beta(t)U_\tau(t),
\end{align}
where $U_\tau(t)=\exp(-2i\tau H_1(t))$ denotes the unitary evolution with $H_1(t)=\sum_x \kappa_{x,t}c^\dag_x c_{x+1}+H.C.$ and $U_\beta(t)=\exp(-2\beta H_2(t))$ denotes the imaginary evolution governed by a random onsite potential $H_2(t)=\sum_x \lambda_{x,t}c^\dag_x c_{x}$. Both $\kappa_{x,t}$ and $\lambda_{x,t}$ are random in space and time with the following simple distribution:
\begin{align}
    P(\kappa_{x,t})&=\frac{1}{2}\delta(\kappa_{x,t}-1)+\frac{1}{2}\delta(\kappa_{x,t}+1)\nonumber\\
    P(\lambda_{x,t})&=\frac{1}{2}\delta(\lambda_{x,t}-1)+\frac{1}{2}\delta(\lambda_{x,t}).
\end{align}
Under this non-unitary random evolution, the wave function evolves as
\begin{align}
    |\psi(T)\rangle=\frac{U}{\sqrt{Z}}|\psi(0)\rangle,\quad \text{with}\ \ Z=\langle \psi(0)|U^\dag U|\psi(0)\rangle.
\end{align}
It is shown in Ref.~\onlinecite{Chen_2020} that this non-unitary dynamics has an emergent two dimensional conformal symmetry for any arbitrary $\beta>0$. The steady state $|\psi(T\to\infty)\rangle$ is critical and has an entanglement entropy which scales logarithmically in the subsystem size, the same as for ground states of critical systems.

If we start with an initial pure Gaussian state, under the non-unitary evolution, $|\psi(T)\rangle$ remains a pure Gaussian state and can be simply written as a product state in some suitable basis, i.e.,
\begin{align}
    |\psi(T)\rangle=\prod_{n=1}^Nf_n^\dag(T)|0\rangle,
\end{align}
where $N$ is the total number of the fermions and is conserved under the time evolution. $\{f_n^\dag(T)\}$ with $n=1,\cdots N$ form a canonical basis for fermion creation operators at time $T$ and satisfy $\{f_m^\dag,f_n^\dag\}=0$. They can be expanded in the  $\{c_x^\dag\}$ basis defined in the spatial direction, i.e.,
\begin{align}
    f_n^\dag=\sum_{x=1}^L u_xc_x^\dag,
\end{align}
where $L$ is the system size and $u_x$ satisfies the normalization constraint $\sum_x|u_x|^2=1$.

\begin{figure*}[t]
\centering
\subfigure[]{\label{fig:tau_fermion} \includegraphics[width=.9\columnwidth]{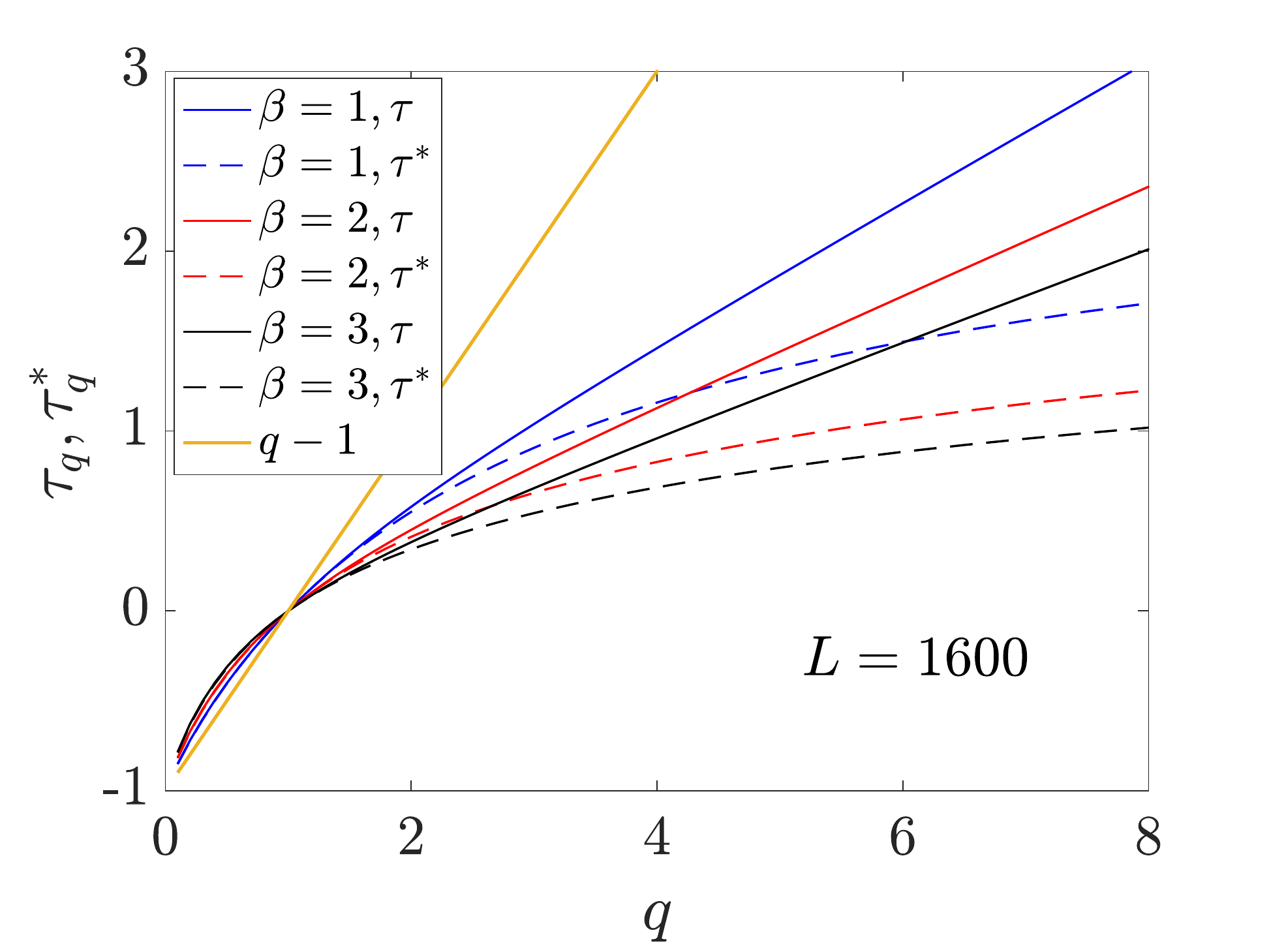}}
\subfigure[]{\label{fig:D2_fermion} \includegraphics[width=.9\columnwidth]{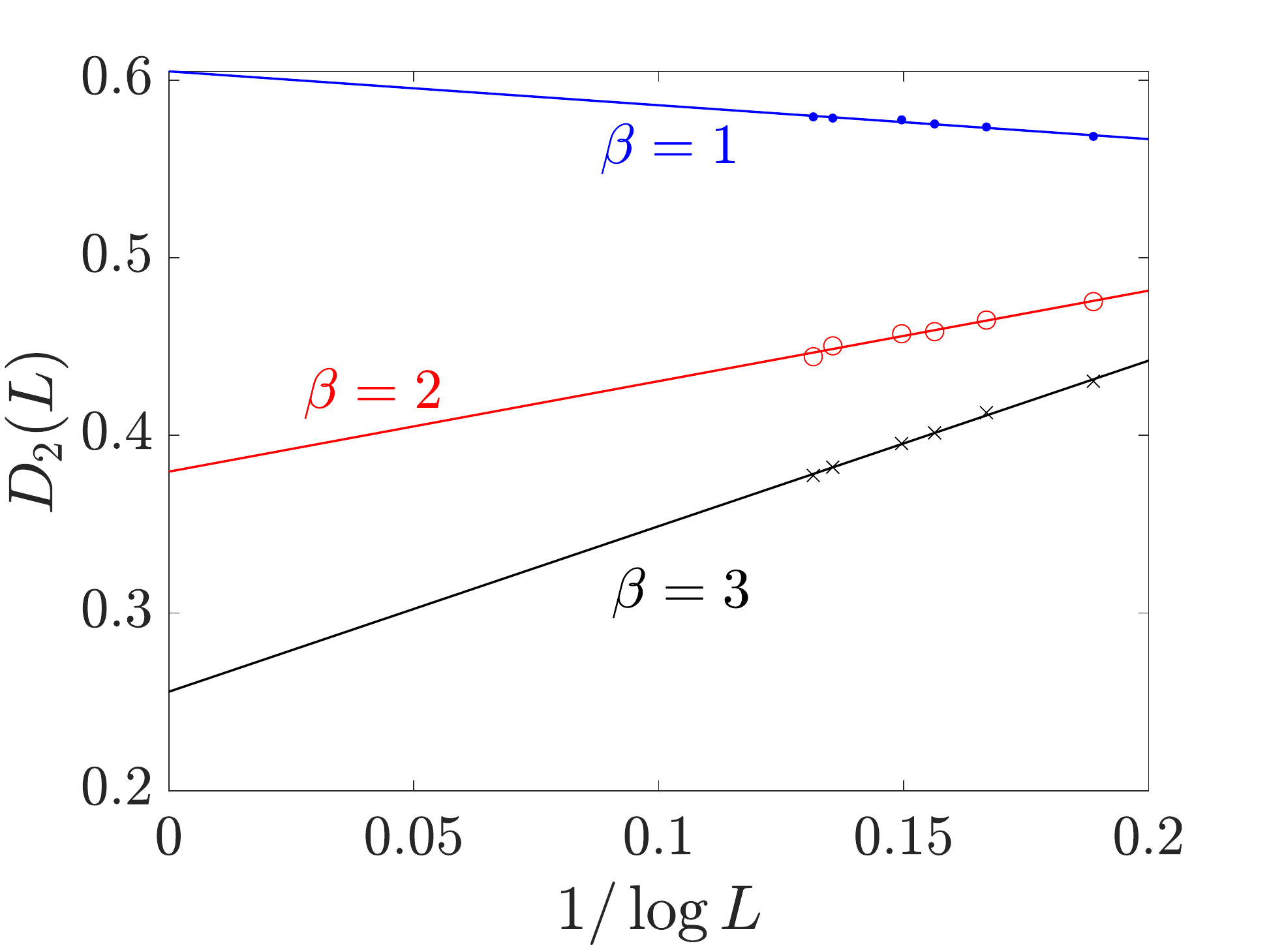}}
\subfigure[]{\label{fig:D_beta} \includegraphics[width=.9\columnwidth]{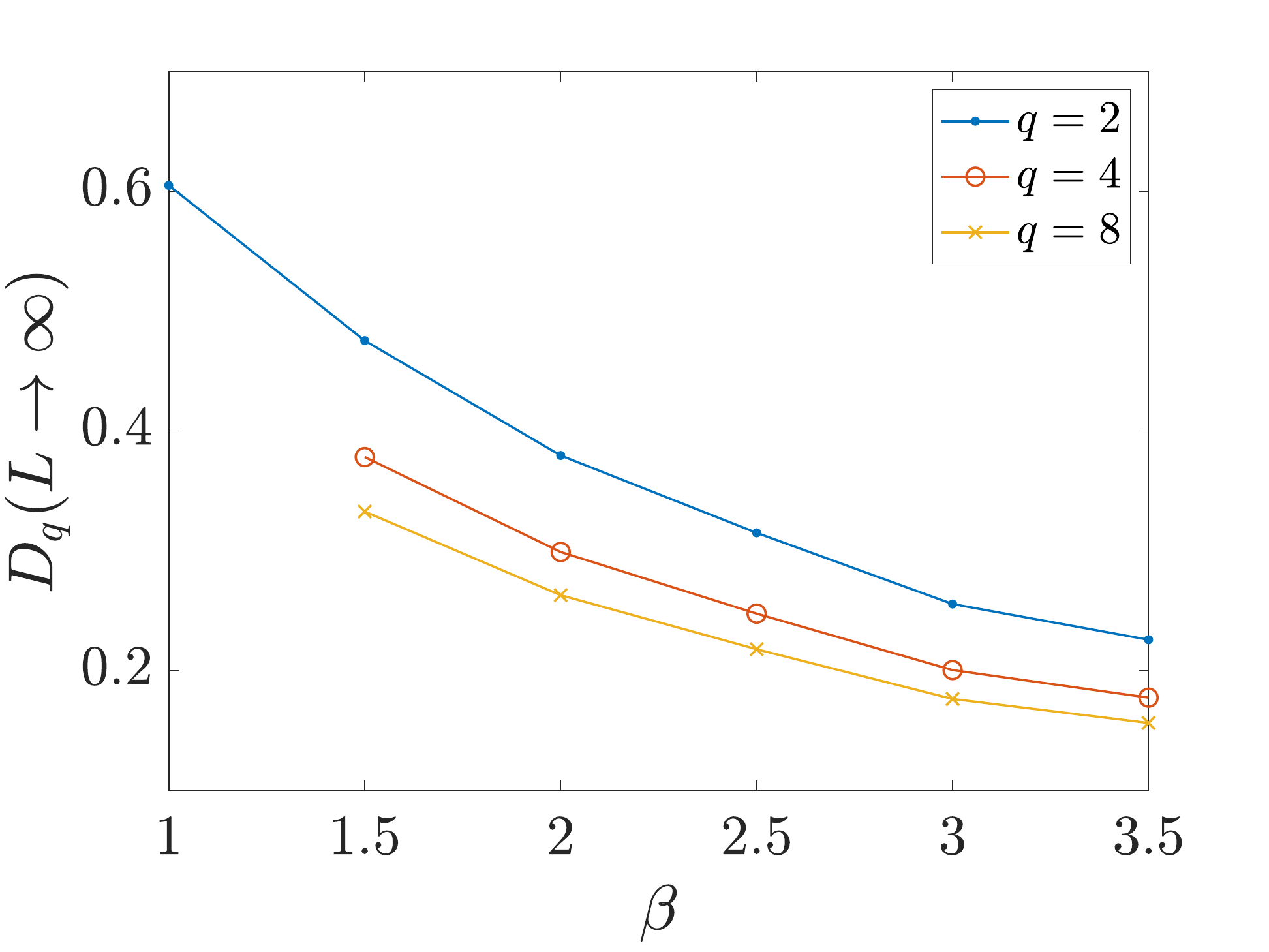}}
\subfigure[]{\label{fig:delta_q} \includegraphics[width=.9\columnwidth]{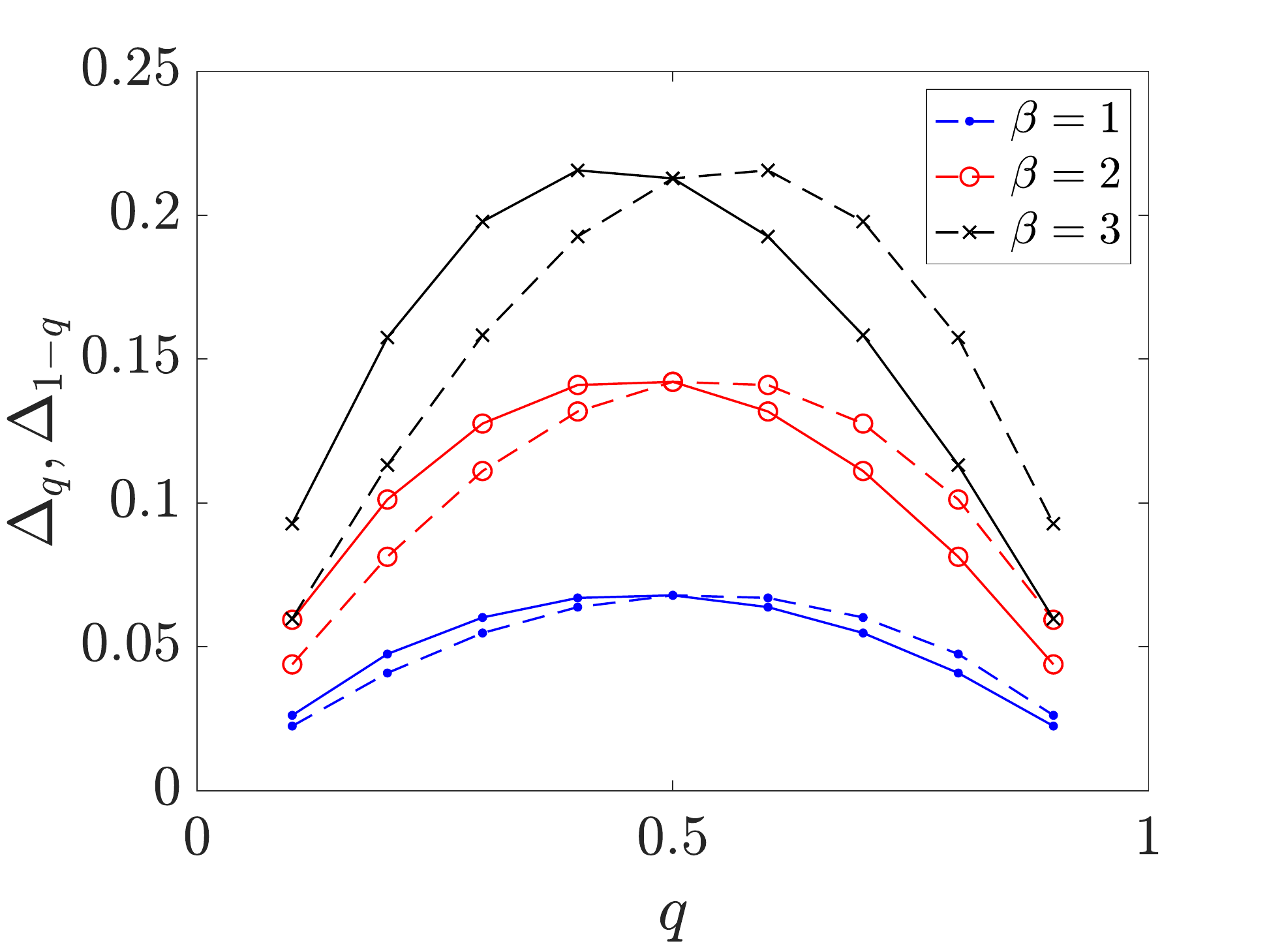}}
\subfigure[]{\label{fig:parabolic} \includegraphics[width=.9\columnwidth]{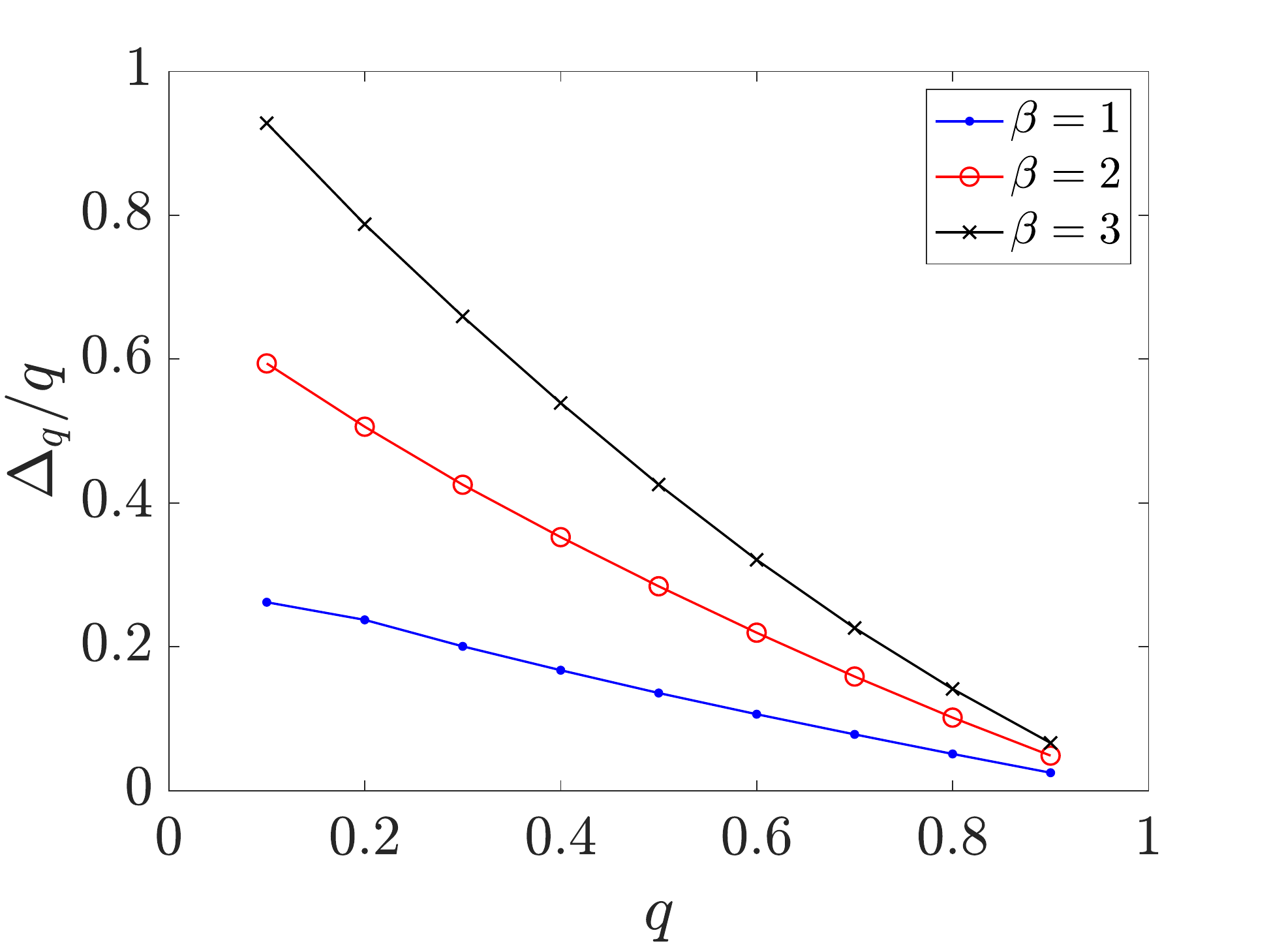}}
\caption{(a) $\tau_q$ and $\tau_q^*$ in the non-unitary free fermion dynamics at various $\beta$. The solid yellow curve is a guideline with $\tau_q=q-1$. (b) $D_2(L)$ vs $1/\log L$ at various $\beta$. The crossing of the curves with the vertical axis gives $D_2(L\to\infty)$. (c) $D_q(L\to\infty)$ as a function $\beta$ for various $q$. (d) The anomalous dimension $\Delta_q$ and $\Delta_{1-q}$ as a function of $q$ at various $\beta$. The solid curve is for $\Delta_q$ and the dashed curve is for $\Delta_{1-q}$. (e) $\Delta_q/q$ vs $q$ at various $\beta$. The curve with $\beta=1$ is close to a straight line.} 
\label{fig:IPR_fermion}
\end{figure*}

We study the spatial distribution of $f_n^\dag$ by computing the q-th moment of the wave function defined as
\begin{eqnarray}
I_q(L) = \sum_{x=1}^L |u_x|^{2q}\sim L^{-\tau_q}.
\end{eqnarray}
As we have done for the Clifford circuit, we compute both $\tau_q$ and $\tau_q^*$ for the steady state and we find that they are different when $q>q_c$, indicating the nonergodicity of the single particle wave function (See Fig.~\ref{fig:IPR_fermion} (a)). Numerically, we find that $q_c$ depends on $\beta$ and decreases as we increase $\beta$. We also observe that $\tau_q$ has a non-trivial dependence on $q$. When $q$ is large, $\tau_q=\alpha q$ with $\alpha<1$, while for small $q$, $\tau_q$ is non-linear in $q$. We further analyze the finite size effect in $\tau_q$ and present the numerical results for $D_2(L)$ in Fig.~\ref{fig:IPR_fermion} (b). Again, we observe that $D_2(L)$ has a logarithmic correction at finite $L$ and  slowly converges to $D_2(L\to\infty)$, which is between 0 and 1 and depends on the value of $\beta$. Similar results are obtained for other $q>1$ and are presented in Fig.~\ref{fig:IPR_fermion} (c). In the unitary evolution limit $\beta=0$, we expect that $D_q\to 1$ and the single particle wave function is uniform in space. On the other hand, in the limit $\beta=\infty$, $D_q=0$ for $q>0$ and the wave function is localized in the space. At finite $\beta$, $D_q$ interpolates between 0 and 1 and forms a set of continuous exponents depending on the parameter $q$, implying a multifractal structure of the wave function.

In addition, we also compute $\tau_q$ in the limit $L\to\infty$ in the regime $0<q<1$ by using the same extrapolation method.  We numerically extract $\tau_q(L\to\infty)$ and present the result for $\Delta_q$ in Fig.~\ref{fig:IPR_fermion} (d). It is approximately symmetric around $1/2$.  In particular, we observe that $\Delta_q\approx \gamma q(1-q)$  when $\gamma$ is small (See the $\beta=1$ curve in Fig.~\ref{fig:IPR_fermion} (e)), similar to what we have observed for the Clifford circuit. 

The non-trivial dependence of $\tau_q$ on $q$ implies that this critical wave function has strong spatial variation and is distinct from the critical state in the free fermion model without disorder, in which the single particle wave function is extended and has a uniform distribution in space. The criticality in these clean systems is caused by the delicate quantum coherence which is fragile and can be easily destroyed when randomness is introduced.  

\section{Discussion and Conclusion}
\label{sec:conclusion}

We investigated the multifractal behavior in two classes of non-unitary random dynamics by numerically examining the inverse participation ratio. In the hybrid random Clifford circuit, we transform the steady state wave function to a graph state characterized by an adjacency matrix. We compute the eigenstates of the adjacency matrix and observe multifractal behavior in the graph space in the non-thermal volume law phase. We further obtain similar multifractal behavior in the volume law phase of the hybrid Clifford QA circuit.
We expect that the multifractality can also be observed in the volume law phase of other random non-unitary Clifford circuit with discrete symmetry \cite{Lavasani_2021,Sang_2021} or generated with only measurement gates \cite{Ippoliti_2021}.

On the other hand, in the non-unitary random free fermion dynamics, the steady state is critical and can be written as $\prod_n f^\dag_1\cdots f^\dag_N|0\rangle$ in some suitable basis. We numerically confirm that these single particle wave functions have strong fluctuations and are multifractal in real space. Previously, the multifractal exponents have been analytically computed in various disordered free fermion systems, including the two dimensional Dirac fermion in a random potential\cite{Chamon,Castillo,Gruzberg} and the power-law  random banded matrix \cite{Evers_RMP}. It would be interesting to generalize and apply these techniques to the non-unitary dynamics in order to analytically compute multifractal exponents, perhaps in some large N non-unitary models\cite{part_one}. 

In the non-unitary free fermion dynamics, the random imaginary potential can be replaced by a continuous weak measurement\cite{alberton2020trajectory,Cao_Tilloy_2019}. In such hybrid dynamics, when the measurement strength is small, the weak measurement is analogous to a random imaginary potential and leads to a similar critical phase with multifractality. As we further increase the measurement rate, a phase transition to an area law entangled phase occurs. The fermions are now fully localized due to the measurement and there is no longer multifractality. 

In a broad sense, the phase transitions in both the hybrid Clifford circuit and the hybrid free fermion circuit with weak measurement are ``Anderson localization"-like. In the former, there is a localization transition in the associated graph space while in the latter, the localization occurs in real space. In both models, before it enters into the Anderson localized phase, there exists a phase in which the wave function has strong fluctuations and is multifractal in nature. This is different from  conventional Anderson localization, where multifractality appears only at the critical point. Recently, some disordered free fermion models exhibiting Anderson localization transitions have been constructed, in which a non-ergodic metallic phase with multifractality is identified, similar to what we have found in this paper \cite{De_Luca_2014,Kravtsov_2015}.  
In particular, the multifractal behavior observed in both the non-unitary Clifford and free fermion models belong to weak mulitfractality class, characterized by a linear growth of $\tau_q$ with large $q$ and an approximate parabolic form in small $q$ \cite{Evers_RMP,Derrida_REM}. It would be interesting to discover non-unitary random dynamics with strong multifractal behavior in which $\tau_q$ becomes zero above $q>q_c$. 

It is widely believed that there is a generic measurement driven transition in an interacting system which  has neither a stabilizer representation nor can be described by a simple free fermion dynamics. We expect that this non-thermal volume law phase has strong random fluctuations and is still multifractal in nature. However, we are  unaware of any good approach to characterize the multifractality correctly and it is also unclear if we can map the measurement induced transition to an Anderson localization transition in a proper basis. We leave these interesting problems for future study.

{\it Note added:} During the completion of this work, we became aware of a work investigating multifractality in the hyrbid Haar random circuit at the critical point in a different context \cite{zabalo2021operator}.
\acknowledgements
 This work is partially supported (J.I.) by the National Science Foundation under Grant Number 1734006. J.I. is supported by a Simons Investigator Award to Leo Radzihovsky from the Simons Foundation. This work is performed in part (X.C.) at Aspen Center for Physics, which is supported by National Science Foundation grant PHY-1607611.

\bibliography{ref}

\end{document}